\documentclass[a4paper,11pt]{article}

\usepackage{jheppub}
\usepackage[usenames,dvipsnames,svgnames,table]{xcolor}
\usepackage{placeins}
\usepackage{booktabs}
\usepackage{ragged2e}
\usepackage{etoolbox}
\usepackage{multirow}
\usepackage{amssymb}
\usepackage{bbold}
\usepackage{booktabs,colortbl}
\usepackage{subfig}
\usepackage{graphicx}
\usepackage[normalem]{ulem}
\usepackage{url}
\usepackage{physics}
\usepackage{makecell}
\usepackage{comment}
\usepackage{multirow}
\apptocmd{\thebibliography}{\justifying}{}{}
\usepackage{array}
\usepackage{hyperref}

\makeatletter\g@addto@macro\bfseries{\boldmath}\makeatother

\definecolor{myblue}{rgb}{0.152941176,0.549019608,0.670588235}

\hypersetup{
    unicode=false,          
    pdftoolbar=true,        
    pdfmenubar=true,        
    pdffitwindow=false,     
    pdfstartview={FitH},    
    pdftitle={Charming Sensitivities for New Higgses at the LHC},
    pdfauthor={Artemis Sofia Giannakopoulou, Patrick Meade, Mauro Valli},
    pdfkeywords={2HDM} {LHC} {Spontaneous Flavor Violation},
    pdfnewwindow=true,
    colorlinks=true,
    linkcolor=myblue,
    citecolor=myblue,
    filecolor=myblue,
    urlcolor=myblue,
    linktocpage=true
}

\def\equationautorefname~#1\null{Eq.\,(#1)\null}
\newcommand{\appendixref}[1]{\hyperref[#1]{appendix~\ref{#1}}}

\definecolor{rosso}{cmyk}{0,1,1,0.4}
\definecolor{rossos}{cmyk}{0,1,1,0.55}
\definecolor{rossoc}{cmyk}{0,1,1,0.2}
\definecolor{blu}{cmyk}{1,1,0,0.3}
\definecolor{blus}{cmyk}{1,1,0,0.6}
\definecolor{bluc}{cmyk}{1,1,0,0.1}
\definecolor{verde}{cmyk}{0.92,0,0.59,0.25}
\definecolor{verdec}{cmyk}{0.92,0,0.59,0.15}
\definecolor{verdes}{cmyk}{0.92,0,0.59,0.4}
\definecolor{bviolet}{rgb}{0.54, 0.17, 0.89}
\definecolor{myred}{rgb}{0.545,0.004,0}

\newcommand{\beq}{\begin{equation}} 
\newcommand{\eeq}{\end{equation}}
\newcommand{\bea}{\begin{eqnarray}}  
\newcommand{\eea}{\end{eqnarray}}
\newcommand{\beastar}{\begin{eqnarray*}}  
\newcommand{\eeastar}{\end{eqnarray*}}
  
\newcommand{\GeV}{\,{\rm GeV}}

\definecolor{Gray}{gray}{0.9}

\title{\boldmath How charming can the Higgs be?}

\author[a]{Artemis Sofia Giannakopoulou,}
\author[a]{Patrick Meade}
\author[b]{and Mauro Valli}

\affiliation[a]{C.N. Yang Institute for Theoretical Physics, Stony Brook University, Stony Brook, NY 11794,~USA}
\affiliation[b]{INFN Sezione di Roma, Piazzale Aldo Moro 2, I-00185 Rome, Italy}

\emailAdd{artemissofia.giannakopoulou@stonybrook.edu}
\emailAdd{patrick.meade@stonybrook.edu}
\emailAdd{mauro.valli@roma1.infn.it}

\abstract{
The coupling of the Higgs boson to first and second generation fermions has yet to be measured experimentally.  There still could be very large deviations in these couplings, as the origin of flavor is completely unknown.  Nevertheless, if Yukawa couplings are modified, especially for light generations, there are generically strong constraints from flavor-changing neutral currents (FCNCs).  Therefore, it is imperative to understand whether there exists viable UV physics consistent with current data that motivates future Higgs coupling probes. In particular, the charm-quark Yukawa is the next quark coupling that could be measured at the LHC {\em if} it is a few times larger than the SM and compatible with flavor data.  This is difficult to achieve in the context of standard ansatz such as Minimal Flavor Violation. In this paper we show that within the framework of Spontaneous Flavor Violation (SFV), using a Two Higgs Doublet Model as an example, the Higgs can be sufficiently charming that new LHC probes are relevant.  In this charming region, we show that new Higgs states near the EW scale with large couplings to quarks are required, providing complementary observables or new constraints on the SM Yukawa couplings. The down-type SFV mechanism enabling the suppression of FCNCs also allows for independent modifications to the up-quark Yukawa coupling, which we explore in detail as well.
} 
\begin{document}
\maketitle
\flushbottom


\section{Overview}
\label{sec:intro}

The study of Flavor Changing Neutral Currents (FCNCs) and CP violation in the Standard Model (SM) naively points to a scale of New Physics (NP) orders of magnitude greater than the electroweak (EW) one, beyond the reach of current and future colliders~\cite{UTfit:2007eik,Isidori:2010kg}. From the perspective of Effective Field Theories (EFT), while the resolution of the SM flavor puzzle might be postponed to a very high ultraviolet (UV) scale, the SM renormalization group is still expected to induce, in general, effects of the Minimal Flavor Violation (MFV) type at low energy~\cite{Buras:2000dm,DAmbrosio:2002vsn}. 
The MFV ansatz allows maximal conservation of the $U(3)^5$ flavor group: as a consequence, the phenomenological impact of indirect searches is minimized~\cite{Silvestrini:2018dos,Aebischer:2020dsw,Aoude:2020dwv}, leaving room for a NP scale within the sensitivity of direct searches~\cite{Grunwald:2023nli,Greljo:2023adz,Garosi:2023yxg,Allwicher:2023shc,Bartocci:2023nvp}. This result undoubtedly makes the MFV framework very popular among Beyond the SM (BSM) practitioners~\cite{Isidori:2019pae,Altmannshofer:2022aml,Glioti:2024hye}.
On the other hand, a MFV scenario corresponds to a highly non-generic situation: the UV theory must be indeed protected against any additional spurion over the Yukawa couplings which break the $U(3)^5$ flavor group in the SM. Notice that this statement must non-trivially translate into selection rules applying also to the CP-violating sector of the BSM scenario. In fact, as a result of SM precision tests such as the unitarity triangle, new $\mathcal{O}(1)$ phases over the one in the Cabibbo-Kobayashi-Maskawa (CKM) matrix can easily push the scale of NP above $100$~TeV~\cite{Bona:2024Ec}. 

An unavoidable outcome of inheriting the hierarchies of the SM Yukawa couplings is that all MFV models feature NP couplings to the third generation as the main phenomenological driver for BSM searches at the Large Hadron Collider (LHC). While this fact might align well with arguments in favor of the EW hierarchy problem~\cite{Giudice:2017pzm,Craig:2022eqo}, the MFV ansatz automatically induces a strong bias into the set of interesting experimental targets at stake for the present and future High Energy frontier. In the absence of hints for top partners or other BSM physics, it is crucial today more than ever to explore alternatives to the MFV paradigm.  This is especially true in the Higgs precision measurements of Yukawa couplings and the search for evidence of yet unmeasured ones.

Observation of Yukawa couplings gives the first direct probe into the origin of known flavor within the SM. However, this direct probe is experimentally very challenging at the LHC.  Reliable flavor identification outside of the leptonic sector is mostly restricted to heavy quarks.  Furthermore, the small size of first and second generation lepton Yukawa couplings means that despite the experimental ability for particle identification, current LHC datasets are not sufficient for measurements.  In particular, only the coupling of the Higgs to the third generation of matter has been experimentally observed thus far~\cite{Dawson:2022zbb}.  Future running of the LHC and its extension, the HL-LHC, should enable the observation of the SM coupling of the Higgs to the muon. However, even with the HL-LHC, measuring the first- or second-generation couplings of the quarks to the Higgs remains extremely challenging. 

The next closest ``observation" of flavor in the quark sector -- and therefore the most tantalizing target at present -- is certainly the charm-quark Yukawa.  This has motivated many recent studies from ATLAS and CMS~\cite{ATLAS:2022ers, ATLAS:2022rej,ATLAS:10years, CMS:2022psv,CMS:2019hve,CMS:10years,ATLAS:2018xfc,ATLAS:2022fnp, ATLAS:2022qef, CMS-PAS-HIG-23-011, ATLAS-CONF-2024-010, CMS-PAS-HIG-23-010, CMS:2022fsq, ATLAS:2024ext}, as well as current HL-LHC projections, showing that it may be possible to probe within a factor of a few times the SM charm Yukawa coupling~\cite{Dawson:2022zbb}. This has also led to a large amount of theoretical activity exploring a variety of observables that potentially enhance the sensitivity of the LHC, some of which have been incorporated into the aforementioned results. These include better attempts at flavor tagging and understanding the exclusive radiative decays into charmed mesons~\cite{Bodwin:2013gca}, looking for associated charm production~\cite{Brivio:2015fxa}, incorporating the sensitivity of the $p_T$ spectrum of the Higgs~\cite{Soreq:2016rae} or the correlated $h+\gamma$ production~\cite{Aguilar-Saavedra:2020rgo}, and combinations of techniques and channels~\cite{Delaunay:2013pja,Perez:2015aoa,Walker:2022yml,Nir:2024oor,Dong:2024gts} (see also references therein).  As of now, even with the concerted theoretical effort and the most advanced experimental analyses based on machine-learning algorithms, the best chance for an observation in this channel seems to require an $\mathcal{O}(1)$ enhancement of the charm Yukawa, putting it well beyond the natural regime of MFV.  If we consider typical models of flavor beyond the SM, it is often difficult to enhance the charm Yukawa significantly without running into other constraints, see e.g., the case study in Ref.~\cite{Glioti:2024hye}. Therefore, it is imperative to explore alternatives to understand if there are motivated possibilities that can be explored by the LHC.  One such framework that was previously investigated for down-type quark Yukawa couplings is Spontaneous Flavor Violation (SFV)~\cite{SFV-first}.

SFV can be viewed as symmetry protected subset of Aligned Flavor Violation (AFV). AFV is achieved by introducing spurions of the maximal SM flavor group as an expansion in powers of the CKM matrix, with expansion coefficients diagonal in flavor space to ensure invariance under rephasing of SM fermion fields~\cite{Branco:1996bq,Pich-Penuelas}.  Generically, the individual quark-flavor number symmetries are not sufficient to leave phenomenological room for low-scale flavored new degrees of freedom \cite{Eberhardt:2020dat,Karan:2023kyj,Bonnefoy:2024gca,SFV-first}. SFV allows for low-scale flavored BSM physics by postulating a CP and family symmetry that is only broken via wave-function renormalization of right-handed up quarks {\em or} down quarks~\cite{SFV-first}. This symmetry breaking pattern allows for enough protection to be phenomenologically viable from indirect constraints while opening up new opportunities at colliders.  In practice one can roughly view SFV as having a new diagonal BSM spurion for {\em either} up-type {\em or} down-type quark couplings, as there are up-type or down-type SFV models\footnote{This refers to sector where symmetry breaking occurs, while the new BSM spurion resides in the opposite quark sector.}. Interestingly, such theories can also enjoy further phenomenological motivation in correspondence to the resolution of the strong CP problem~\cite{Nelson:1983zb,Barr:1984qx,Bento:1991ez,Hiller:2001qg}.

If one applies SFV to a Two Higgs Doublet Model (2HDM), this extends 2HDMs beyond the typical four discrete choices implied by Natural Flavor Conservation and allows for EW scale states that couple with fundamentally different patterns to quark flavors.  In~\cite{Egana-Ugrinovic:2019dqu,Egana-Ugrinovic:2021uew} the up-type version of a SFV 2HDM was studied and, in particular, if mixing is allowed, it lead to significant enhancements in the down and strange Yukawa couplings of the SM Higgs.  In the context of a flavor model, this also allows for the correlation of the SM Higgs Yukawa with additional observables.  For instance, a mixing between Higgses in combination with new Yukawas of a heavy second Higgs implies a new type of di-Higgs signature~\cite{Egana-Ugrinovic:2021uew}, which for the down and strange Yukawa couplings sets strongest bounds.  

In this paper we extend the analysis to the SFV down-type 2HDM where significant enhancements of the charm Yukawa are possible. This is crucial for two main reasons. First, it is important to determine if there are perturbatively describable models compatible with indirect flavor bounds that the LHC could detect through a charm Yukawa measurement. Second, in models that permit a significant enough deviation in the charm Yukawa coupling to be measurable, we need to assess whether this measurement is the optimal observable. Alternatively, we must consider if the relevant parameter space for charm Yukawa sensitivity is already excluded by other LHC observables.  

Our main finding here is that in the context of SFV 2HDM models there is a viable parameter space where a charm Yukawa could be observable at the LHC while satisfying all current bounds.  Futhermore, as with up-type SFV models~\cite{Egana-Ugrinovic:2021uew} multiple observables beyond charm specific ones may be as sensitive as single-Higgs properties. Therefore, a key lesson from our study is that a concerted effort across searches and precision measurements is recommended to extract the maximal LHC reach for BSM physics.

The paper is organized as follows. In \autoref{sec:The model} we introduce the class of 2HDMs realizing SFV. In \autoref{sec:flavor_bounds} and \autoref{sec:ColliderBounds} we discuss and derive the main results of our work, namely the constraints on the parameter space of the down-type SFV 2HDM, coming from flavor and collider physics respectively.  Additional indirect constraints such as EW precision tests are subleading since 2HDM models naturally preserve a custodial symmetry~\cite{HiggsHunters}, and as discussed in \autoref{sec:The model} we work in the CP-conserving version of the model for simplicity.  We then compare with the results from Higgs precision measurements in ~\autoref{sec:precision} to understand how complementary these measurement are, and whether there is relevant parameter space that can be explored with flavor tagging measurements at the LHC. Finally, in Appendix \ref{sec:updated_up_type} we update the bounds on up-type SFV 2HDM and we leave our conclusions to \autoref{sec:concl}. For the sake of completeness, in Appendices \ref{loop_functions} and \ref{couplings} we report the loop functions relevant to this study and collect the couplings of the Higgs particles to SM fermions.

\section{Spontaneous Flavor Violating 2HDM}
\label{sec:The model}

\qquad Following the SFV prescription described in \autoref{sec:intro}, in this study we work with a specific class of models realizing it, namely 2HDMs endowed with the SFV ansatz. We illustrate how well-motivated BSM benchmarks that are not flavorless or mostly interacting via the third-generation fermions can be probed by experiments. 

The Higgs sector of a 2HDM is extended compared to the Standard Model by adding a second Higgs doublet with the same hypercharge as the SM one. It is convenient to work in the Higgs basis \cite{Davidson:2005cw}, where only one of the two doublets has a non-zero vacuum expectation value. Choosing $\langle H_1 \rangle \neq 0$ and $\langle H_2 \rangle = 0$, we can write:
\begin{equation}
        H_1 =\begin{pmatrix}
                G^+ \\
                \frac{1}{\sqrt{2}} \left( v + H_1^0 +iG^0 \right) 
               \end{pmatrix}, 
        \qquad
        H_2 = \begin{pmatrix}
                    H^+ \\
                    \frac{1}{\sqrt{2}} \left( H_2^0 + iA \right)
                \end{pmatrix}.
    \label{eq:doublets}
\end{equation}
Here we denote the Goldstone bosons by $G^0$ and $G^\pm$. In this extended scalar sector we have five mass eigenstates. We will denote by $H^{\pm}$ the charged Higgs eigenstate, $A$ the CP-odd one,  $H_1^0$ and $H_2^0$ the fields mixing into the CP-even light and heavy neutral Higgs:
\begin{equation}
    \begin{split}
        h &= H_1^0 \sin \left( \beta - \alpha \right) + H_2^0 \cos \left( \beta - \alpha \right) ,  \\
        H &= H_1^0 \cos \left( \beta - \alpha \right) - H_2^0 \sin \left( \beta - \alpha \right), \\
    \end{split}
    \label{eq:HiggsCPeven}
\end{equation}
where $\cos \left(\beta - \alpha \right)$ is the alignment parameter. According to what was reported above, in what follows we shall identify $h$ with the $125 \, \GeV$ Higgs resonance observed at the LHC. 

The Higgs sector of the Lagrangian of the models considered in this work is given by:
\begin{equation}
    |D_{\mu}H_a|^2 - V \left( H_1, H_2 \right) - \left( \mathcal{Y}^u_{aij} \overline{Q}_{Li} H_a U_{Rj} + \mathcal{Y}^{d}_{aij} \overline{Q}_{Li} H_a^c D_{Rj} + \mathcal{Y}^{\ell}_{aij} \overline{L}_{Li} H_a^c \ell_{Rj} + h.c. \right) \ , \
    \label{eq:HiggsLagrangian}
\end{equation}
where we denote with $\mathcal{Y}^f_{aij}$ the Yukawa matrices of our model, with $f=u,d,\ell$ denoting up-type, down-type quarks and leptons respectively. The index $a=1,2$ runs over the two distinct Higgs doublets. The potential $V\left( H_1 , H_2 \right)$ in the Higgs basis reads as:
\begin{equation}
    \begin{split}
        V \left( H_1 , H_2 \right) &= m_1^2 H_1^{\dagger} H_1 +m_2^2 H_2^{\dagger} H_2 + \left( m_{12}^2 H_1^{\dagger}H_2 + h.c. \right) + \frac{1}{2} \lambda_1 \left( H_1^{\dagger} H_1 \right)^2 \\
        &+ \frac{1}{2} \lambda_2 \left( H_2^{\dagger} H_2 \right)^2 + \lambda_3 \left( H_1^{\dagger} H_1 \right) \left( H_2^{\dagger} H_2 \right) + \lambda_4 \left( H_1^{\dagger} H_2 \right) \left( H_2^{\dagger} H_1 \right) \\
        &+ \left\{ \frac{1}{2} \lambda_5 \left( H_1^{\dagger} H_2 \right)^2 + \left[ \lambda_6 \left( H_1^{\dagger} H_1 \right) +  \lambda_7  \left( H_2^{\dagger} H_2 \right) \right] \left( H_1^{\dagger} H_2 \right) + h.c. \right\}.
    \end{split}
    \label{eq:Higgspotential}
\end{equation}
The alignment angle related to the physical Higgs states can be expressed via the parameters appearing in Eq.~\eqref{eq:Higgspotential}, namely:
\begin{equation}
    \tan \left[ 2 \left( \beta - \alpha \right) \right] = \frac{2 \lambda_6 v^2}{\lambda_1 v^2 - \left( m_2^2 + \frac{1}{2} \left( \lambda_3 + \lambda_4 + \lambda_5 \right) v^2 \right)}.
    \label{eq:tanbma}
\end{equation}
In the limit $\cos \left( \beta - \alpha \right) \to 0 $, the SM Higgs field $h$ resides entirely in one of the two Higgs doublets. This regime is called the \textit{aligned limit} of the theory. We can tune the parameters of our theory such that the model is in this regime by setting in Eq.~\eqref{eq:tanbma} $m_2 \to \infty$ \textit{(decoupling limit)} or $\lambda_6 \to 0$ \textit{(alignment without decoupling)}. 

In our study we work with $\cos \left( \beta - \alpha \right) \neq 0$, i.e. assuming both terms are present in each line of Eq.~\eqref{eq:HiggsCPeven}. Indeed, when the alignment parameter is non-zero, deviations from the SM predictions for the Higgs couplings are expected and can be interestingly probed by current measurements \cite{Craig:2013}. In this work, we are interested in exploring the small but non-zero $\cos \left( \beta -\alpha \right)$, such that we remain relatively close to the SM. Additionally, for simplicity, we assume that the new Higgs eigenstates are degenerate, i.e., $\lambda_4, \lambda_5 \to 0$, such that $m_H = m_{H^\pm} = m_A$. We will maintain this assumption of degenerate masses throughout the rest of this analysis. Under these conditions, we can write the following expansion for the alignment parameter \cite{Egana-Ugrinovic:2015vgy, GunionHaber:2HDM}:
\begin{equation}
    \cos \left( \beta - \alpha \right) = - \left| \lambda_6 \right| \frac{v^2}{m_H^2} + \mathcal{O}\left( \frac{v^4}{m_H^4} \right),
    \label{eq:cosbma}
\end{equation}
where $\lambda_6$ is given in equation (\ref{eq:Higgspotential}). Assuming a small but non-zero mixing between the two Higgs doublets, we can use Eq.~\eqref{eq:cosbma} to estimate the range of masses, $m_H$, for which the coupling $\lambda_6$ remains finite. To avoid Landau poles at the TeV scale, we set an upper limit of $m_H \leq 2 \, \rm TeV$. Within this mass range and for a small alignment parameter, $\lambda_6$ stays within the perturbative regime \cite{Egana-Ugrinovic:2019dqu}. 

For the sake of minimality, in the present analysis we do not introduce any additional phases apart from the CKM one, working with a CP-conserving SFV 2HDM. Also, notice that without loss of generality we can always rotate the Higgs potential in Eq.~\eqref{eq:Higgspotential} to the Peccei-Quinn basis where the couplings and masses are real.

As anticipated in \autoref{sec:intro}, an important characterization for the phenomenology of the SFV 2HDM is whether the alignment in flavor space goes along with the SM down-type or up-type Yukawa matrix. For instance, in the case of alignment with the down-quark sector, the up-quark fields will need to be rotated to the mass eigenstates according to:
\begin{equation}
    Q = \begin{pmatrix}
            V^\dagger u \\
            d
        \end{pmatrix} \ ,
    \label{eq:redefinition_of_Q} 
\end{equation}
where we denote by $V$ the CKM matrix. Following the spurion language of \cite{SFV-first}, for the \textit{down-type} SFV, there will be no new flavor violating spurions transforming under $U(3)_Q \times U(3)_{d}$ appearing in renormalizable interactions. However, new flavor aligned spurions non-trivial under $U(3)_Q \times U(3)_{u}$ will be allowed. In such a case, the couplings to the Higgs doublet $H_1$ will read as:
\begin{equation}
\mathcal{Y}_1^u = V^* {\rm diag}(y_u,y_c,y_t)  \equiv V^* Y^u \, ,  \quad  \mathcal{Y}_1^{d} = - Y^d,  \quad \mathcal{Y}_1^{\ell} = - Y^{\ell} \ ,
\label{eq:YukawasH1}
\end{equation}
while Yukawa matrices dictating the couplings of the second Higgs doublet $H_2$ will be:
\begin{equation}
\mathcal{Y}_2^u = V^* {\rm diag} \left( \lambda_u, \lambda_c, \lambda_t \right) \equiv  V^* \Lambda^u,  \quad \mathcal{Y}_2^{d} = -\xi Y^d,  \quad  \mathcal{Y}_2^{\ell} = -\xi^{\ell} Y^{\ell} \ ,
\label{eq:YukawasH2}
\end{equation}
where we denote by $Y^f$ the (diagonal) SM Yukawa matrices with $f=u,d, \ell$ depending on the coupling to up- or down-type quarks or to leptons. Using Eq.s~\eqref{eq:YukawasH1}~--~\eqref{eq:YukawasH2}, we can derive the couplings of the Higgs eigenstates to the quarks and leptons, getting for example for the lightest CP even Higgs,
\begin{equation}
\mathcal{Y}^h_{u_i \overline{u}_j}  = \delta_{ij} \left[ y_{u_j} \sin \left( \beta - \alpha \right) + \lambda_{u_j} \cos \left( \beta - \alpha \right) \right] \ . 
\label{eq:Yukawa2HDM}
\end{equation}
The rest of the Higgs mass eigenstates couplings are listed in Appendix \ref{couplings}. Note that $\lambda_u, \lambda_c,\lambda_t$ are  new, independent Yukawa couplings of $H_2$ to the up-type quarks.\footnote{Not to be confused with the couplings appearing in Eq.~\eqref{eq:Higgspotential}, denoted by $\lambda_i$ with $i=1,\dots,6$.}. We emphasize that these BSM couplings are not bound to follow the SM hierarchy and the second Higgs doublet $H_2$ is allowed to couple to the quarks with different coupling size than the SM analogue.
In Eq.~\eqref{eq:YukawasH2}, $\xi$ and $\xi^{\ell}$ are constants that control the size of the couplings of the second Higgs doublet to the down-type quarks and leptons, respectively.\footnote{In our analysis we study in detail the quark sector and we set for simplicity $\xi^{\ell}=0$.} As mentioned before, they are required to be real since we don't allow for new sources of CP violation. Finally, couplings to gauge bosons follow the standard literature on 2HDM~\cite{GunionHaber:2HDM}.

In a similar fashion, one can define the case of an \textit{up-type} SFV 2HDM where new Higgs couplings to the down-quark sector are not forbidden a priori and will impact the phenomenology \cite{Egana-Ugrinovic:2019dqu}. While we will update also the available bounds for that scenario in Appendix \ref{sec:updated_up_type}, the primary focus of this study will be on the constraining power of the charm coupling modifier $\kappa_c$, according to the definition:
\begin{equation}
 \kappa_{q_i} \equiv \mathcal{Y}_{q_i \bar{q}_i}^h \big/ y_{q_i} \,,
 \label{eq:kappa}
\end{equation} 
where we denote with $y_{q_i}$ the coupling of the Higgs within the SM to the quark $q$ with flavor $i$, and with $\mathcal{Y}_{q_i \bar{q}_i}^h$ the corresponding one in the mass basis of the 2HDM, as explicitly reported in Eq. \eqref{eq:Yukawa2HDM}. Therefore, we will primarily explore the \textit{down-type} SFV 2HDM scenario. In the following, we will first present the most relevant bounds from  and then move to inspect the constraints coming from direct searches at colliders. 

\section{Bounds from FCNC processes}
\label{sec:flavor_bounds}
\qquad Rare $B$ decays and neutral meson oscillations are precisely measured FCNC processes, which can provide crucial tests of the SM. As in the SM, in SFV theories FCNCs do not appear at tree level. Nevertheless, also for the class of models analyzed here $\Delta F = 1,2$ FCNC transitions offer the most relevant constraints for what concerns indirect searches \cite{Egana-Ugrinovic:2019dqu}. Using an EFT approach, the relevant Wilson coefficients for the generated dimension-six operators can be computed for the SFV 2HDM under scrutiny. Adopting up-to-date measurements and theory predictions of the most recent literature -- reported in \autoref{tab:flavor_summary} -- we can restrict the available parameter space of the new Yukawa couplings $\lambda_u, \ \lambda_c, \ \lambda_t$.

\begin{table}
    \centering
    \renewcommand{\arraystretch}{2}
    \begin{tabular}{|c|c|c|c|}
         \hline
         \textbf{Observable} & \textbf{Constraint} & \textbf{95\% probability range} & \textbf{Ref.} \\
         \hline
         $b \to s  \gamma$ & \makecell{ $C_7$ \\ $C^{\prime}_7$ }& $ \makecell{ \left[-0.062,0.040 \right] \\ \left[-0.30,0.30 \right] } $ & \cite{Misiak:2020vlo, HFLAV:2022esi} \\
         \hline
         $b \to d  \gamma$ & \makecell{ $C_7$ \\ $C^{\prime}_7$ }& $ \makecell{ \left[-0.11, 0.20 \right] \\ \left[-0.41 , 0.41 \right] } $ & \cite{Bause:2022rrs,BaBar:2010vgu} \\
         \hline
        $b \to s \, \mu^+ \mu^-$ &  $C_{10}$ & $\left[ -0.15, 0.62 \right]$ & \cite{Ciuchini:Bsmumu}\\
         \hline
        $K - \overline{K}$ & \makecell{$Re(C_1)$ \\ $Im(C_1)$ } & \makecell{ $\left[ -6.8, 7.7 \right] \times 10^{-13}$ GeV$^{-2}$\\ $\left[-1.2, 2.4 \right] \times 10^{-15}$ GeV$^{-2}$} & \cite{Silvestrini:2018dos,Bona:2024Ec} \\
        \hline
        $B_s - \overline{B}_s$ & $|C_1|$  & $< 1.9 \times 10^{-11}$ GeV$^{-2}$ & \cite{Silvestrini:2018dos,Bona:2024Ec} \\
        \hline
        $B_d - \overline{B}_d$ & $|C_1|$  & $< 9.5 \times 10^{-13}$ GeV$^{-2}$ & \cite{Silvestrini:2018dos,Bona:2024Ec} \\
        \hline
        $D - \overline{D}$ & \makecell{ $Re(C_1)$ \\ $Im(C_1)$}  & \makecell{ $\left[-2.5, 3.1 \right] \times 10^{-13}$ GeV$^{-2}$ \\ $\left[-9.4, 8.9 \right] \times 10^{-15}$ GeV$^{-2}$}& \cite{Silvestrini:2018dos,Bona:2024Ec}  \\
        \hline
    \end{tabular}
    \caption{Most important constraints in our analysis from indirect searches. We list the available limits on the Wilson coefficients at a matching scale close to the EW one. Notice that $C_7^{\prime}$ and $D - \overline{D}$ bounds are relevant only for up-type SFV studied in the Appendix \ref{sec:updated_up_type}.}
    \label{tab:flavor_summary}
\end{table}
 
\subsection{Bounds from rare \boldmath$B$ decays}
\label{sec:Rad_b_decays}
\begin{figure}[t!]
    \centering
    \includegraphics[scale=1.5]{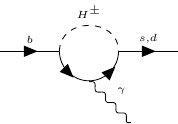}
    \caption{Example of penguin diagram for SFV 2HDM contributing to inclusive radiative $B$ decays. Notice that non-negligible effects from the fermion in the loop can arise a priori from any up-type quark on the basis of the size of the new Yukawa couplings of the theory.}
    \label{diagram:btosgamma} 
\end{figure}

\qquad The inclusive radiative $B$-meson decays are well-known to provide leading constraints on the parameter space of 2HDMs \cite{Hermann:2012fc}. In our class of 2HDMs, $B \rightarrow X_{s,d} \, \gamma$  arise at one loop as shown in \autoref{diagram:btosgamma}: NP contributions to these processes can be mediated by either charged or neutral Higgs eigenstates. In the case of down-type SFV, working with an operator product expansion at mass dimension six, the effective weak Hamiltonian involves the same set of operators of the SM one \cite{Crivellin:2HDM}; for $b \to s \gamma$, e.g.,  we can write:
\begin{equation}
    \mathcal{H}_{eff}^{b \to s \gamma} = -\frac{4G_F}{\sqrt{2}} V_{tb} V_{ts}^* \sum_i C_i \mathcal{O}_i \ ,
    \label{eq:bsgamma_ham}
\end{equation}
where the sum is indeed over current-current, QCD and QED penguin and dipole operators. In particular, the leading phenomenological contributions to the process from the down-type SFV 2HDM are given by the matching onto the electromagnetic and chromomagnetic dipole operators that are not chirally suppressed by the light-quark mass, namely:
\begin{align}
    \mathcal{O}_7 &= \frac{e}{16\pi^2}m_b  \, \overline{s} \sigma^{\mu \nu} P_R b \, F_{\mu \nu} \ , \nonumber \\ 
    \mathcal{O}_8 &= \frac{g_s}{16 \pi^2} m_b  \, \overline{s} \sigma^{\mu \nu} T^a P_R b \, G^a_{\mu \nu} \ .
    \label{O7}
\end{align}
The Wilson coefficients corresponding to the operators in Eq.~\eqref{O7} can be explicitly calculated for our model \cite{Crivellin:2HDM} using the couplings in Appendix \ref{couplings}. They are presented below as functions of the relevant SM and BSM parameters:
\begin{equation}
       C_7 = \frac{v^2}{V_{tb}V_{ts}^*} \sum_{j=u,c,t} \left( \frac{1}{m_b} \mathcal{Y}^{H^+ *}_{sj} \mathcal{Y}^{H-}_{jb} \frac{\mathcal{C}^0_{7,XY}(z_j)}{m_j} +  \mathcal{Y}^{H^+*}_{sj} \mathcal{Y}^{H^+}_{bj} \frac{\mathcal{C}^0_{7,YY}(z_j)}{m_j^2} \right) \ ,
    \label{Wils7}
\end{equation}
where $z_j \equiv m^2_j / m^2_H$ and the loop functions are reported for convenience in Appendix \ref{loop_functions}. We can obtain the expressions for the Wilson coefficients of $\mathcal{O}_8$ by replacing the loop functions in Eq.~\eqref{Wils7} by $\mathcal{C}^0_{8,XY}$ and $\mathcal{C}^0_{8,YY}$, also reported in the same appendix. Eventually, the same loop functions also characterize the matching expression for the dipole operators with opposite chirality with respect to the ones in Eq.~\eqref{Wils7} -- which matter for the up-type SFV 2HDM -- implying also a trivial change in the Yukawa couplings involved \cite{Crivellin:penguins}.

Using the latest experimental average for the branching ratio of the inclusive decay $B \to X_s \gamma$ -- namely $\mathcal{B}_{s\gamma} = (3.49 \pm 0.19) \times 10^{-4}$~\cite{HFLAV:2022esi} -- and adopting the theoretical prediction updated in Ref.~\cite{Misiak:2020vlo}, we derive a bound on the 2HDM parameter space $(m_H,\lambda_i)$. Similarly, following what was recently worked out in Ref.~\cite{Bause:2022rrs} for $b \to d \gamma$, we investigate the bound from $B \to X_d \gamma$ in the same parameter space, adopting $\mathcal{B}_{d\gamma} = (14.1 \pm 5.7) \times 10^{-6}$ as the experimental constraint \cite{BaBar:2010vgu}. 
With no much of a surprise, the bound reported in \autoref{tab:flavor_summary} highlights how the leading constraint comes from $b \to s \gamma$, whose theory prediction with respect to $b \to d \gamma$ is less sensitive to long-distance effects \cite{Crivellin:penguins}. In particular, we find the driver of the bound to be the electromagnetic dipole, whose contribution to the branching ratio is a factor of four more relevant than the chromomagnetic contribution under QED $\otimes$ QCD  renormalization group, see the general phenomenological expressions given in \cite{Hurth:2003dk} and the updated one for $b \to s \gamma$ in  \cite{Misiak:2020vlo}, where NP effects on the branching ratio are specifically reported for $\mathcal{O}_{7,8}$.

\begin{figure}[t!]
    \centering
    \includegraphics[scale=1.5]{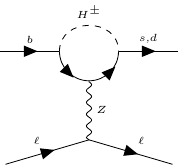}
    \caption{Example of penguin diagram for SFV 2HDM contributing to the short-distance physics of $b \to s,\mu^+ \mu^-$. Notice that the loop function of these contributions features an important suppression in the limit of very light quarks in the loop \cite{Inami:1980fz}.}
    \label{diagram:btosmumu}
\end{figure}

In addition to the radiative inclusive channel, a rare $B$-meson decay which provides today a precise indirect test of the short-distance physics of the SM is certainly $B_{s} \to \mu^+ \mu^-$ \cite{Czaja:2024the}, particularly well-studied in the context of 2HDMs \cite{Logan:2000iv,Lang:2022mxu}. 
Such a decay pertains to the partonic process $b \to s \mu^+ \mu^-$: within our flavor construction, tree-level diagrams that violate flavor are not allowed and the leading-order contributions to this class of FCNC transitions come at the one loop from Z penguins and new box diagrams involving charged Higgses, see \autoref{diagram:btosmumu}. At dimension six, the effective Hamiltonian is the one given in Eq.~\eqref{eq:bsgamma_ham} with the addition of the semileptonic operators:
\begin{equation}
    \begin{split}
        \mathcal{O}_9 & =  \frac{e^2}{16\pi^2}\overline{s}\gamma_{\rho}P_L b \, \overline{\mu} \gamma^{\rho} \mu \ , \\
        \mathcal{O}_{10} & = \frac{e^2}{16\pi^2} \overline{s}\gamma_{\rho}P_L b \, \overline{\mu} \gamma^{\rho} \gamma_5 \mu \ ,
    \end{split}
    \label{O9&10}
\end{equation}
and their chirality-flipped version, which turn out to play a minor role in our study. Due to the quantum numbers characterizing $\mathcal{O}_9$, sizable hadronic physics can plague a precise assessment of the short-distance contribution associated to this operator in a global analysis of $b \to s \, \ell^+ \ell^-$ transitions \cite{Ciuchini:2021smi,Bordone:2024hui}. As a consequence of this observation, a conservative bound on NP in $C_9$ points inevitably to a relatively weak constraint \cite{Ciuchini:Bsmumu}. 

On the contrary, the absence of $\gamma$ penguins in the effective coupling associated to $\mathcal{O}_{10}$ underlies a robust probe of new short-distant effects, experimentally well constrained by the competitive measurement of the branching ratio for $B_s \to \mu^+ \mu^-$ performed independently by ATLAS, CMS and LHCb collaborations~\cite{HFLAV:2022esi}. Using the couplings in Appendix \ref{couplings} and the results collected in Ref.~\cite{Crivellin:bsmumu}, we can write down the explicit expression for the Wilson coefficient of $\mathcal{O}_{10}$ in our SFV 2HDM:
\begin{equation}
        C _{10}  =  \frac{\mathcal{Y}^{H^+*}_{s q} \mathcal{Y}^{H^+}_{bq}}{2e^2 V_{qb}V_{qs}^*} \left( I_1( z_q) -1 \right) \, , 
    \label{Wils10}
\end{equation}
where $I_1(z_i)$ is a loop function given in Appendix \ref{loop_functions} and $z_q = m_q^2 / m_H^2$ with ${q=u,c,t}$. Notably, such a function vanishes in the limit of massless quarks, implying the relevance of the top quark running in the loop for the $B_s \to \mu^+ \mu^-$ in our model-building scenarios. 

Using the bounds from rare $B$ decays, we can map the allowed parameter space for the new Yukawa couplings of the second Higgs doublet. The results are presented in \autoref{fig:combined_flavor}. The bounds are shown in terms of the mass of the heavy Higgs and the relevant new Yukawa coupling. For each panel, we chose to allow only one of the up-type Yukawas of the second Higgs doublet to be non-zero. This particular slicing of the parameter space is chosen in order to make the presentation of the results clear and highlight the role of these bounds for each of the three up-quark flavors. 

The shape of the constraints depends on the couplings of the second Higgs doublet to the fermions. Modifying the proportionality constant $\xi$ that relates the new down-type Yukawa matrix to the SM one,  leads to different results for the bounds presented in \autoref{fig:combined_flavor}. In that figure, we have chosen to set $\xi = 0$ for the sake of simplicity, bearing in mind that $\mathcal{Y}^{H^-} = - (\xi V^* Y^d) \neq 0$ would result in more Wilson coefficients becoming relevant. Furthermore, notice that the CKM insertions and quark masses multiplying the loop functions in the matching expression for the dipole operators are different when we choose to have a non-zero charm or top new Yukawa coupling. In this regard, for the case of the up-quark Yukawa, the loop functions in combination with the CKM elements that will appear in the expressions lead to a large suppression of the overall Wilson coefficient which consequently does not lead to interesting constraints on the parameter space.  

For the case of the semi-leptonic operators entering in the effective Hamiltonian for $b \rightarrow s \mu^+ \mu^-$, Eq.~\eqref{Wils10}, there is no term $ \propto \left( \mathcal{Y}^{H^+} \mathcal{Y}^{H^-} \right)$. As a result, there is no dependence on $\xi$ for the bound from $B_s \to \mu^+\mu^-$.  Finally, notice also that since it is the charged Higgs eigenstate $H^{\pm}$ which mediates these FCNC processes, meaning that the result will be independent of the alignment parameter $\cos ( \beta - \alpha )$, as evident from what reported in Appendix \ref{couplings}. 

\subsection{Bounds from neutral meson mixing}
\label{sec:Neutral_meson_mixing}
\begin{figure}[t!]
    \centering
    \subfloat[]{{\includegraphics[scale=1.7]{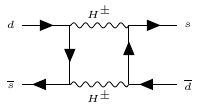} }}%
    \subfloat[]{{\includegraphics[scale=1.7]{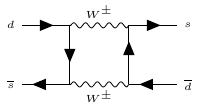} }}%
    \caption{Example diagrams for $K-\overline K$ mixing. On the left we show the contribution mediated by the new charged Higgs eigenstates, next to the $W$-boson box diagram of the SM on the right. For the other $\Delta F = 2$ processes examined one obtains the relevant diagrams by properly changing the quark flavors of the initial and final states.}%
    \label{diagram:boxes}%
\end{figure}
\qquad Meson-antimeson mixing offers the most stringent bounds on baryon- and lepton-number conserving NP \cite{Silvestrini:2018dos,Aebischer:2020dsw}. Continuing our analysis, we will now consider the constraints on the new Yukawas from the bounds on the Wilson coefficients of the $\Delta F = 2$ effective operators responsible of those FCNC processes. 

The relevant Feynman diagrams are at the one-loop level, mediated by charged Higgs eigenstates or charged EW bosons, see the examples in \autoref{diagram:boxes}. Then, the $\Delta F = 2$ effective weak Hamiltonian describing neutral meson mixing reads in full generality \cite{UTfit:2007eik}:
\begin{equation}
    \mathcal{H}^{\Delta F =2}_{eff} = \sum_{i=1}^{5} {C}_i\mathcal{O}_i + \sum_{j=1}^{3} {C}_{j}^{\,\prime}\mathcal{O}_{j}^{\,\prime} + h.c.\ ,
\end{equation}
where, e.g., in the case of $K - \overline{K}$  mixing, the effective operators are:
\begin{equation}
    \begin{split}
        \mathcal{O}_1 &= \left(\overline{s}_{\alpha} \gamma_{\mu} P_L d_{\alpha} \right) \left( \overline{s}_{\beta} \gamma^{\mu} P_L d_{\beta} \right), \\ 
        \mathcal{O}_2 &= \left(\overline{s}_{\alpha} P_Ld_{\alpha} \right) \left( \overline{s}_{\beta} P_L d_{\beta} \right),\\ 
        \mathcal{O}_3 &= \left( \overline{s}_{\alpha} P_L d_{\beta} \right) \left( \overline{s}_{\beta} P_L d_{\alpha} \right), \\ 
        \mathcal{O}_4 &= \left( \overline{s}_{\alpha} P_L d_{\alpha} \right) \left( \overline{s}_{\beta} P_R d_{\beta} \right), \\ 
        \mathcal{O}_5 &= \left( \overline{s}_{\alpha} P_L d_{\beta} \right) \left( \overline{s}_{\beta} P_R d_{\alpha} \right) , \\
        \mathcal{O}_1^{\,\prime} &= \left(\overline{s}_{\alpha} \gamma_{\mu} P_R d_{\alpha} \right) \left( \overline{s}_{\beta} \gamma^{\mu} P_R d_{\beta} \right), \\ 
        \mathcal{O}_2^{\,\prime} &= \left(\overline{s}_{\alpha} P_R d_{\alpha} \right) \left( \overline{s}_{\beta} P_R d_{\beta} \right),\\ 
        \mathcal{O}_3^{\,\prime} &= \left( \overline{s}_{\alpha} P_R d_{\beta} \right) \left( \overline{s}_{\beta} P_R d_{\alpha} \right), 
    \end{split}
    \label{DF2O}
\end{equation}
with $P_{L,R} = \frac{1}{2}(1\mp \gamma_5)$ and $\alpha$, $\beta$ are color indices. Notice that for $D - \overline{D}$ and $B_{d,s }- \overline{B}_{d,s }$ meson mixing, the four-fermion operators of the corresponding $\Delta F = 2$ weak effective Hamiltonian need the appropriate change in the field flavor content in Eq.~\eqref{DF2O}. 

The corresponding one-loop Wilson coefficients are presented below in terms of the relevant masses, mixing angles and Yukawa couplings. The loop functions $D_0$ and $D_2$ involved in the $\Delta F = 2$ matching are reported in Appendix \ref{loop_functions}. 
The Wilson coefficients for box diagrams with charged Higgs in the loops are \cite{Crivellin:2HDM}:
\begin{equation}
    \begin{split}
        C_1 &= \frac{-1}{128 \pi^2} \sum_{j,k=1}^{3} \mathcal{Y}_{d_1 \overline{u}_j}^{H^+ *} \mathcal{Y}_{d_2 \overline{u}_j}^{H^+} \mathcal{Y}_{d_1 \overline{u}_k}^{H^+*} \mathcal{Y}_{d_2 \overline{u}_k}^{H^+} D_2(m_{u_j}^2,m_{u_k}^2,m_H^2,m_H^2), \\ 
        C_2 &= \frac{-1}{32\pi^2} \sum_{j,k=1}^{3} m_{u_j} m_{u_k}  \mathcal{Y}_{u_j \overline{d}_1}^{H^- *} \mathcal{Y}_{d_2 \overline{u}_j}^{H^+} \mathcal{Y}_{u_k \overline{d}_1}^{H^- *} \mathcal{Y}_{d_2 \overline{u}_k}^{H^+} D_0(m_{u_j}^2,m_{u_k}^2,m_H^2,m_H^2), \\ 
        C_4 &= \frac{-1}{16 \pi^2} \sum_{j,k=1}^{3} m_{u_j}m_{u_k} \mathcal{Y}_{u_j \overline{d}_1}^{H^- *} \mathcal{Y}_{d_2 \overline{u}_j}^{H^+} \mathcal{Y}_{d_1 \overline{u}_k}^{H^+ *} \mathcal{Y}_{u_k \overline{d}_2}^{H^-} D_0(m_{u_j}^2,m_{u_k}^2,m_H^2,m_H^2), \\ 
        C_5 &= \frac{1}{32 \pi^2} \sum_{j,k=1}^3 \mathcal{Y}_{u_j \overline{d}_1}^{H^- *} \mathcal{Y}_{u_j \overline{d}_2}^{H^-} \mathcal{Y}_{d_1 \overline{u}_k}^{H^+ *} \mathcal{Y}_{d_2 \overline{u}_k}^{H^+} D_2(m_{u_j}^2,m_{u_k}^2,m_H^2,m_H^2) ,
    \end{split}
    \label{boxes_Higgs}
\end{equation}
and by replacing $\mathcal{Y}^{H^+} \leftrightarrow \mathcal{Y}^{H^-}$ we obtain the coefficients of the primed operators. For the processes via charged Higgs and $W$ or Goldstone exchange, one gets \cite{Crivellin:2HDM}:
\begin{equation}
\begin{split}
    C_1 &= \frac{-1}{128 \pi^2} \frac{g^2}{m_W^2} \sum_{j,k =1}^3 V_{j1}^* V_{k2}m_{u_j}m_{u_k} \mathcal{Y}_{d_2 \overline{u}_j}^{H^+} \mathcal{Y}_{d_1 \overline{u}_k}^{H^+*} \bigg[ D_2 ( m_{u_j}^2, m_{u_k}^2, m_W^2,m_H^2 ) \\
    & \hspace{4cm}- 4m_W^2 D_0( m_{u_j}^2, m_{u_k}^2, m_W^2,m_H^2) \bigg], \\
    C_1' &= \frac{-1}{128 \pi^2}\frac{g^2}{m_W^2} \sum_{j,k=1}^3 V_{j1}^* V_{k2}m_s m_d \mathcal{Y}_{u_j \overline{d}_2}^{H^-} \mathcal{Y}_{u_k \overline{d}_1}^{H^-*} D_2(m_{u_j}^2, m_{u_k}^2, m_W^2,m_H^2) \ , \\
    C_2 &= \frac{-1}{32\pi^2}\frac{g^2}{m_W^2} \sum_{j,k=1}^3 V_{j2}V_{k1}^* m_d m_{u_j}^2 m_{u_k} \mathcal{Y}_{u_j \overline{d}_1} ^{H^- *} \mathcal{Y}_{d_2 \overline{u}_k}^{H^+} D_0(m_{u_j}^2, m_{u_k}^2, m_W^2,m_H^2) \ ,\\
    C_2' &= \frac{-1}{32 \pi^2} \frac{g^2}{m_W^2} \sum_{j,k=1}^3 V_{j1}^* V_{k2} m_s m_{u_j}^2 m_{u_k} \mathcal{Y}_{u_j \overline{d}_2}^{H^-} \mathcal{Y}_{d_1\overline{u_k}}^{H^+*} D_0(m_{u_j}^2, m_{u_k}^2, m_W^2,m_H^2) \ ,\\    
    C_4 &= \frac{-1}{32 \pi^2} \frac{g^2}{m_W^2} \sum_{j,k=1}^3 \bigg[  V_{j1}^* V_{k2} \big( m_{u_j} m_{u_k} m_d m_s \mathcal{Y}_{d1 \overline{u}_k}^{H^+*} \mathcal{Y}_{d_2 \overline{u}_j}^{H^+} + m_{u_j}^2m_{u_k}^2 \mathcal{Y}_{u_j \overline{d}_2}^{H^-} \mathcal{Y}_{u_k \overline{d}_1}^{H^-*} \big)  \\
    & \times D_0(m_{u_j}^2, m_{u_k}^2, m_W^2,m_H^2) -m_W^2 V_{k1}^* V_{j2} \mathcal{Y}_{u_j \overline{d}_1}^{H^-*} \mathcal{Y}_{u_k \overline{d}_2}^{H^-} D_2(m_{u_j}^2, m_{u_k}^2, m_W^2,m_H^2) \bigg] \ ,\\
    C_5 &= \frac{1}{64 \pi^2} \frac{g^2}{m_W^2} \sum_{j,k=1}^3 \bigg( V_{j1}^*V_{k2}m_{u_j}m_s \mathcal{Y}_{d_2 \overline{u}_j}^{H^+} \mathcal{Y}_{u_k \overline{d}_1}^{H^-*} +V_{j2}V_{k1}^* m_{u_j}m_d \mathcal{Y}_{d_1 \overline{u}_j}^{H^+*} \mathcal{Y}_{u_k \overline{d}_2}^{H^-} \bigg) \\
    & \hspace{4cm} \times D_2(m_{u_j}^2, m_{u_k}^2, m_W^2,m_H^2)  \  .
    \label{boxes_EW}
\end{split}
\end{equation}
While for $B_{d,s }- \overline{B}_{d,s }$ the expressions above hold up to a trivial change of flavor indices, notice that to obtain the Wilson coefficients for the $D - \overline{D}$ mixing one needs to substitute $\mathcal{Y}^{H^+ *} \leftrightarrow \mathcal{Y}^{H^-}$.

To set the constraints on the Yukawa coupling - heavy Higgs mass parameter space, we exploit the results derived by the \textsc{UTfit} collaboration from the unitarity triangle analysis generalized to the case of NP \cite{UTfit:2022hsi}, where only tree-level flavor changing processes are assumed to be SM-like, while FCNC amplitudes are modified to account for generic NP effects. We use the up-to-date bounds on the $\Delta F =2$ Wilson coefficients recently presented in \cite{Bona:2024Ec}. We apply the constraints on the Wilson coefficients directly at a matching scale close to the EW one following the strategy in \cite{Silvestrini:2018dos} (see their Table 1), barring accidental cancellations among different contributions from $\Delta F =2$ operators. 

We present the results for each generation separately by allowing only one new coupling to be non-zero for each panel presented in \autoref{fig:combined_flavor}. Notice that the parameter $\xi$ does not affect the strongest bounds we get from $\Delta F = 2$: unless we move far away from the $\xi \sim 0$ regime, the effect on the resulting constraints is minor. On the other hand, there is no dependence on the value of $\cos (\beta -\alpha)$ due to NP contributions involving charged Higgs states. We observe a small difference between the resulting bounds from $B_d$ and $B_s$ oscillations coming merely from the different CKM elements appearing in the expression for the $\Delta F =2$ matching. The bounds in \autoref{fig:combined_flavor} correspond to the processes providing a constraint which limits the available parameter space in a visible manner in the plot. 

\subsection{Summary of flavor bounds}
\label{sec:summary_flavor}
\begin{figure}[h!]
    \centering
    \includegraphics[scale=0.5]{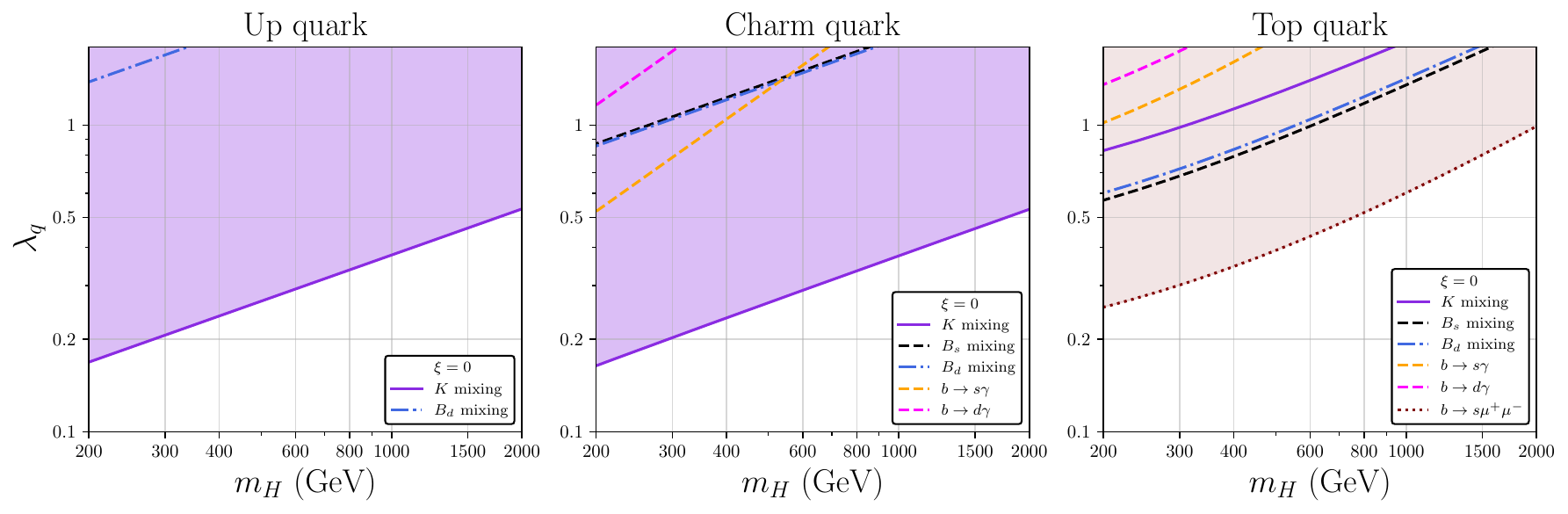}
    \caption{Bounds from FCNCs. Left Panel: Bounds on $\lambda_u$ vs.\ $m_H$ parameter space, setting $\xi=0$ and $\lambda_c=\lambda_t=0$. Middle Panel: Bounds on $\lambda_c$ vs.\ $m_H$ parameter space, setting $\xi=0$ and $\lambda_u=\lambda_t=0$. Right Panel: Bounds on $\lambda_t$ vs. \ $m_H$ parameter space, setting $\lambda_u = \lambda_c = 0$.  These results can be sensitive to the choice of the proportionality constant $\xi$ but are independent from different choices in the value of $\cos (\beta -\alpha)$. In the panels above, the dashed orange lines correspond to the constraints coming from the $b \to s \gamma$ decays and the dotted magenta lines to the ones from $b \to s \mu^+ \mu^-$ processes. For neutral meson mixing, we only display the processes that yield a constraint for values up to $|\lambda_q| \sim 1$. We observe that for the down-type SFV 2HDM the bounds from $D$ meson mixing are not constraining enough to limit the available parameter space in a significant way. Also, notice that $\Delta B = 2$ constraints are bounds on the absolute value of the Wilson coefficients.}%
    \label{fig:combined_flavor}
\end{figure}
\qquad We summarize here the bounds in the plane $\lambda_q$ vs.\ $m_H$ for the SFV down-type 2HDM. In \autoref{sec:Rad_b_decays} and \autoref{sec:Neutral_meson_mixing} we present the details for the constraints we derived using the analysis of dimension-six effective operators contributing to rare $B$ decays and neutral meson mixing. The combined constraints are presented in \autoref{fig:combined_flavor}. In each panel of the figure we allowed only one of the new Yukawas to be non-zero, assuming the other ones to be vanishing.  Notice that for all the panels of \autoref{fig:combined_flavor} we have set $\xi=0$, i.e. we neglected any effect of the coupling of the second Higgs doublet $H_2$ to the down-type quarks. Changing the value of this parameter can potentially lead to a modification of the flavor constraints shown. Also, the loop functions depend on the masses of the virtual states involved, impacting the resulting constraints. For example, in the case of  $b \to s \mu^+ \mu^-$, the limit of the loop function for light-quark masses leads to a suppression of the constraint observed in the top-quark panel of \autoref{fig:combined_flavor}. The differences between the constraints for the charm and up Yukawas can be attributed to the different CKM insertions that are relevant for the Wilson coefficients, as well as the size of the bounds that were used for the coefficients themselves, see table \autoref{tab:flavor_summary}. Notice that for the case of $K$ mixing, the expression for $C_1$ when we set $\lambda_c \neq 0$ is almost identical to the one when $\lambda_u \neq 0$, because the CKM elements that appear are similar. This fact does not hold for $B$ mixing. The constraints we get from $B_{d,s}$ for $\lambda_u$ are consequently not as restrictive as they are for $\lambda_c$. 

Overall, we find that the Flavor Physics constraints allow for viable parameter space where the new Yukawa couplings can be larger than their SM counterparts. For example, looking at the middle panel of \autoref{fig:combined_flavor}, we see that $\vert \lambda_c \vert \sim 0.2 $ is allowed and corresponds to an enhancement of $\mathcal{Y}_{hc\overline{c}}$  about 3.7 times the SM value of $y_c$ for $\cos \left( \beta - \alpha \right) = 0.1$.

\section{Collider Direct Search Constraints}
\label{sec:ColliderBounds}

As shown in~\autoref{sec:flavor_bounds}, flavor constraints still allow for new Higgs states with large couplings to light quarks at the EW scale.  Therefore it is possible to have {\em direct} collider search constraints that are potentially as powerful as the precision flavor probes.  In the limit of perfect alignment $\cos (\beta-\alpha)=0$, the new heavy 2HDM states can be directly produced through both Drell-Yan production as well as quark fusion due to the large possible Yukawa couplings.  Given our simplifying assumption of approximate degeneracy between $H,A$ and $H^\pm$ from \autoref{sec:The model}, the strongest constraints come from the direct production of a single heavy Higgs state that decays into SM particles which dominate the branching fractions.  In the $\cos (\beta-\alpha)=0$ limit, while there are interesting constraints from decays exclusively into dijets~\cite{Egana-Ugrinovic:2019dqu}, there is no $H-h$ mixing and therefore the Yukawa couplings of the SM are unaffected. Since we are interested in understanding the allowed parameter space for large deviations in the SM Yukawa couplings, we explore the small but non-zero $\cos (\beta-\alpha)$ limit.  In this regime, as shown in~\cite{Egana-Ugrinovic:2021uew}, the decays to SM di-bosons and di-Higgs can quickly dominate the dijet branching fractions and have more sensitive cross section constraints to our underlying production mechanism in the relevant mass window.  The dominant collider search constraints come from new heavy Higgs states being singly produced in quark fusion followed by decays such as $H\rightarrow ZZ$, $A\rightarrow Zh$, and $H\rightarrow hh$.  For simplicity we also set the $\xi$ parameters to 0 when investigating the direct search bounds. This does not significantly change the collider phenomenology in the regions allowed by flavor as the large {\em new} quark Yukawa couplings dominate direct tree-level production and the non-zero alignment parameter singles out the di-boson and di-Higgs decays in this region.

\begin{table}[h]
    \centering
    \renewcommand{\arraystretch}{1.5}
    \resizebox{\textwidth}{!}{%
    \begin{tabular}{|c|c|c|c|c|}
        \hline
        \textbf{Process}  & \textbf{Experiment} & \textbf{Reference} & \textbf{$\mathcal{L} (\mathrm{fb}^{-1})$} & \textbf{Search Window [GeV]} \\
        \hline
        
         $H \to ZZ$  & ATLAS & \cite{ATLAS:H-to-ZZ} & 139  & [210,2000] \\
         \hline
         $H \to hh$ & \makecell{ATLAS \\ CMS} & \makecell{\cite{ATLAS:dihiggs-bb}\cite{ATLAS:dihiggs-gaga}\cite{ATLAS:dihiggs-tata} \\ \cite{CMS:H_hh_gagabb}} & \makecell{$139$ \\ $[126,139]$} & [251,5000] \\
         \hline
         $A \to Zh$ & ATLAS & \cite{ATLAS:AZh} & 139 & [220,5000]\\
         \hline
         
    \end{tabular}%
    }
    \caption{Table of direct resonant searches used to constrain the model with results shown in \autoref{fig:LHC_combined}.}
    \label{tab:collider_refs}
\end{table}

\begin{figure}[h]
    \centering
    \subfloat[]{{\includegraphics[scale=0.46]{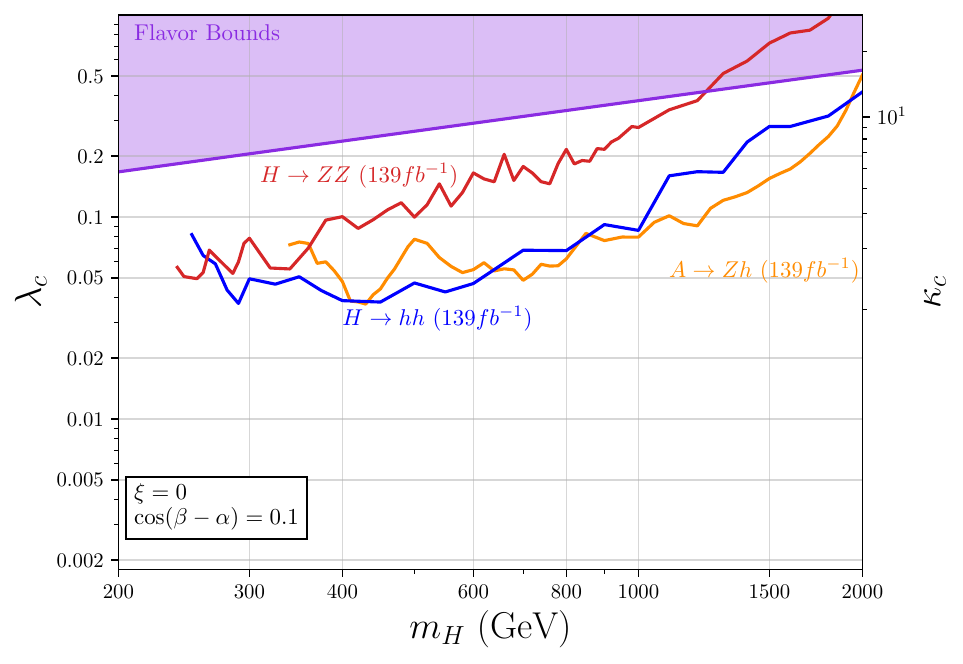} }}%
    \subfloat[]{{\includegraphics[scale=0.46]{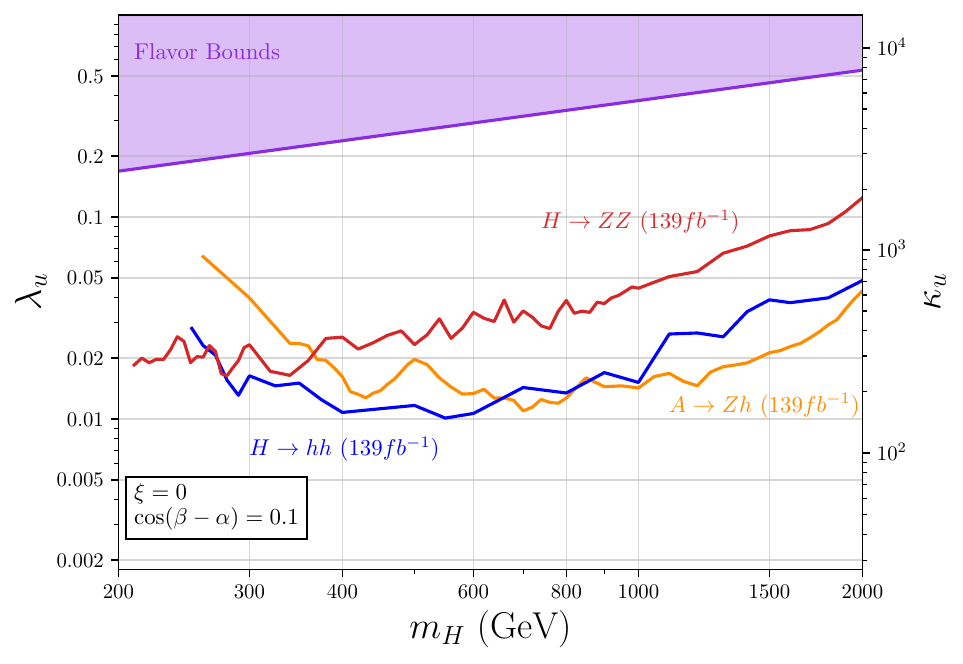} }}\\ 
    \subfloat[]{{\includegraphics[scale=0.46]{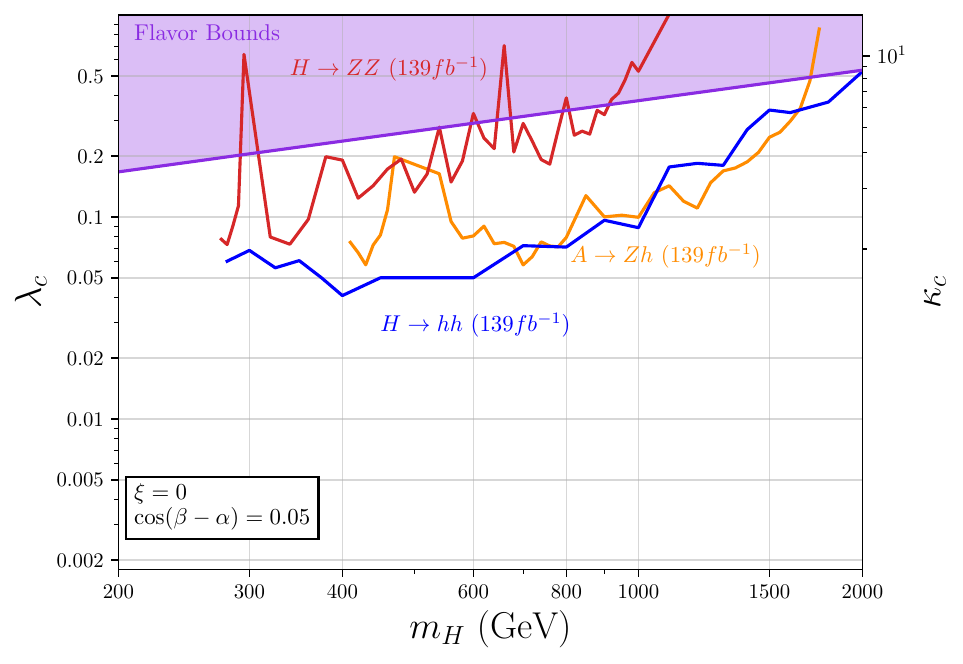} }}%
    \subfloat[]{{\includegraphics[scale=0.46]{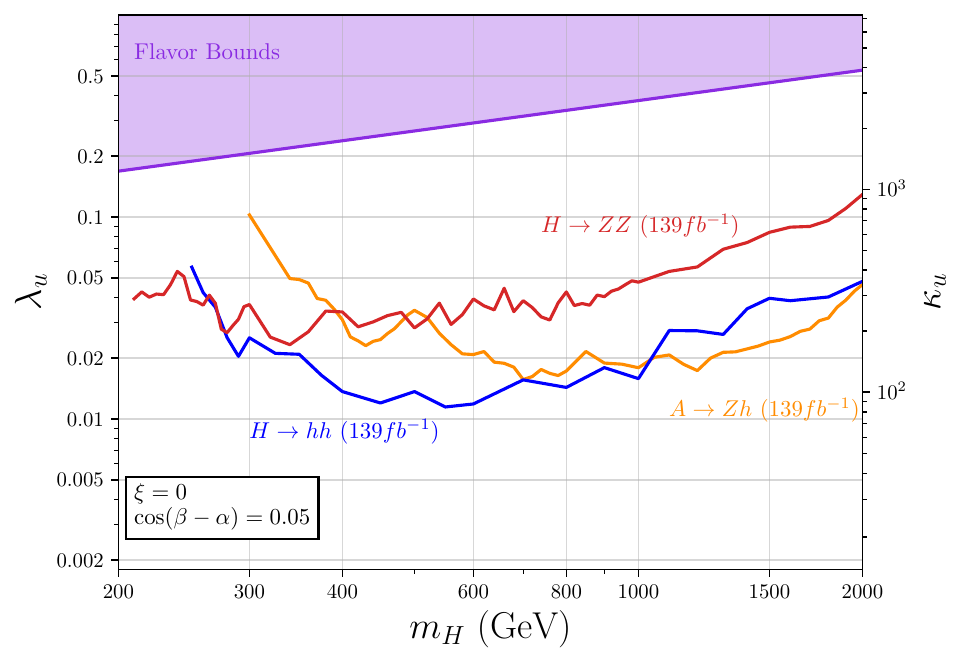} }}%
    \caption{Collider searches for the 2HDM model with couplings to charm in panel (a) and up in quarks panel (b), overlaid with the appropriate flavor bounds derived in \autoref{sec:flavor_bounds}. The couplings of the second doublet to the down-type quarks are set to zero i.e. $\xi=0$ and the alignment parameter is set to $\cos (\beta -\alpha ) =0.1$. The vertical axis on the right-hand side of each panel shows the coupling modifier of the charm and up quarks respectively. The combined flavor bounds are presented in purple. We present the bounds of resonant production of the heavy Higgs H, decaying to different channels. The bounds from decays to a pair of Z bosons \cite{ATLAS:H-to-ZZ} are in red and to a pair of $125 \, \GeV$ Higgs \cite{ATLAS:dihiggs-bb, ATLAS:dihiggs-gaga, ATLAS:dihiggs-tata} are in blue. In orange, we present the bounds from resonant production of the CP odd A decaying to a Z boson and a $125 \, \GeV$ Higgs boson \cite{ATLAS:AZh}. Panels (c) and (d) are the same as (a) and (b) with $\cos \left( \beta - \alpha \right) = 0.05$.}%
    \label{fig:LHC_combined}%
\end{figure}

To determine the bounds from direct searches we use the results from resonant di-Higgs searches \cite{ATLAS:dihiggs-bb, ATLAS:dihiggs-gaga, ATLAS:dihiggs-tata, CMS:H_hh_gagabb, CMS:2024rgy}, $A \to Z h$ \cite{ATLAS:AZh} and $H \to ZZ$ \cite{ATLAS:H-to-ZZ}, whose integrated luminosities and mass regions are summarized in \autoref{tab:collider_refs}.  We recast the cross section bounds from the ATLAS and CMS searches onto our relevant parameter space by simulating the 2HDM in \textsc{MadGraph5} \cite{Madgraph5}. We implemented a 2HDM \textsc{FeynRules} \cite{Feynrules2.0} file, following the SFV prescription outlined earlier. By inputting this model into \textsc{MadGraph5} \cite{Madgraph5}, we were able to scan over the relevant $\lambda_q$ - $m_H$ parameter space for a fixed $\cos (\beta-\alpha)$ to find the direct search constraints.  Given that we are interested in how charming the Higgs can be, as well as the novel enhancement of the up-quark Yukawa, we present the results for $\lambda_c$ - $m_H$ and $\lambda_u$ - $m_H$ for $\cos \left( \beta - \alpha \right) = 0.1$ in panels (a) and (b) of \autoref{fig:LHC_combined}.  From this figure we also see that the direct search constraints are currently stronger than the flavor bounds {\em despite} having a non MFV texture and EW scale Higgs masses.  This emphasizes the power of flavor changing suppression in SFV scenarios as well as the strength of the LHC to directly probe these flavorful states at the EW to TeV scale.   On the right hand axis we also show the modification to the appropriate SM Higgs Yukawa coupling $\kappa_{q_i}$ for the given input parameters of the model.  In panels (b) and (c) of ~\autoref{fig:LHC_combined} we show the same parameter space constraints, but for $\cos \left( \beta - \alpha \right) = 0.05$.  This demonstrates how the $\kappa_{q_i}$ parameter shifts accordingly, but also slight changes in the direct search constraints due to the dependence of the di-boson couplings on the alignment parameter. Note, however that the new Yukawa coupling is unaffected in the small alignment parameter limit.  In all panels of  \autoref{fig:LHC_combined} we see that the bounds on $\lambda_u$ are stronger than those on $\lambda_c$ due to the enhanced PDFs of the up quark compared to the charm quark. However interpreted in terms of the $\kappa_{q_i}$ they are very different due to the underlying SM Yukawa coupling predictions.

\section{Higgs precision and Yukawa Coupling measurements}
\label{sec:precision}

We are now in a position to attempt to answer our titular question. As shown in~\autoref{sec:flavor_bounds} and \autoref{sec:ColliderBounds}, flavor and direct search collider constraints place strong bounds on an down-type SFV 2HDM model.  However, as demonstrated, there is still the possibility that these models can generate large deviations in the up and charm SM Higgs Yukawa couplings. Therefore it is important to correlate with the precision measurements of the SM Higgs couplings to understand how complementary these measurements are.

A common approach for single-Higgs precision at the LHC is based on exclusive final states that are correlated with a production mechanism for the Higgs and reported as signal strength modifiers:
\begin{align}
    \mu_{if} = \frac{\sigma_{i} \times {\rm BF}_f}{\sigma_{i}^{\rm SM} \times {\rm BF}_f^{\rm SM}}.
    \label{mu_kappa}
\end{align}

Large modifications of light quark Yukawas due to our 2HDM can affect the signal strengths in a number of ways.  First, if there is mixing between the heavy and light Higgs states to generate Yukawa coupling modifications there is an inherent dilution of SM Higgs couplings.  Second, in the limit of a large Yukawa modification, the branching fractions to {\em all} exclusive final states will be modified which can be correlated across production mechanisms since the total width changes as
\begin{equation}
    \frac{\Gamma_{tot}^{\rm SM}}{\Gamma_{tot}} = \left[\sin^2(\beta-\alpha) \left(1+\frac{\Gamma_{qq}^{\rm SM}}{\Gamma_{tot}^{\rm SM}} \left( \frac{\lambda_q^2}{\sin^2(\beta-\alpha)y_q^2} -1 \right) \right) \right]^{-1} .
    \label{dilution}
\end{equation}
A correlated effect, but subleading for most measurements, is the modification of the production cross section both in terms of loop effects as well as enhancing additional production modes. Finally, if a light quark Yukawa coupling is large enough, it is possible for the statistics to be sufficient for a direct measurement via flavor tagging. The effects can be quantified through Higgs coupling modifier fits with $\kappa_{q_i}$ or in an EFT framework.  Given that we have the full 2HDM model it is more straightforward to use coupling modifiers to correlate across observables. The explicit dependence on the model parameters for $\kappa_{q_i}$ defined in Eq.~\eqref{eq:kappa} can be found in Appendix \ref{couplings}. For the first two effects on Higgs precision there is no measurement of the modified Yukawa coupling ``directly", therefore the bounds they provide on the up and charm Yukawa couplings are less robust to other possible BSM effects.  Nevertheless, in the context of the specific model we study, they give constraints on both $\kappa_c$ and $\kappa_u$ that we refer to as normalization constraints. As mentioned in~\autoref{sec:intro}, one can go beyond signal strengths and include differential shape effects in the $p_T$ spectrum, but these represent marginal improvements currently.  When the coupling modifier is large enough, additional production modes, exclusive decays and flavor tagging can provide measurements that are easier to directly correlate with the change in the light quark Yukawa couplings.   However, these bounds are typically much weaker, and only promising at the LHC for modifications of the charm Yukawa coupling.

\begin{figure}[h]
    \centering
    \subfloat[]{{\includegraphics[scale=0.46]{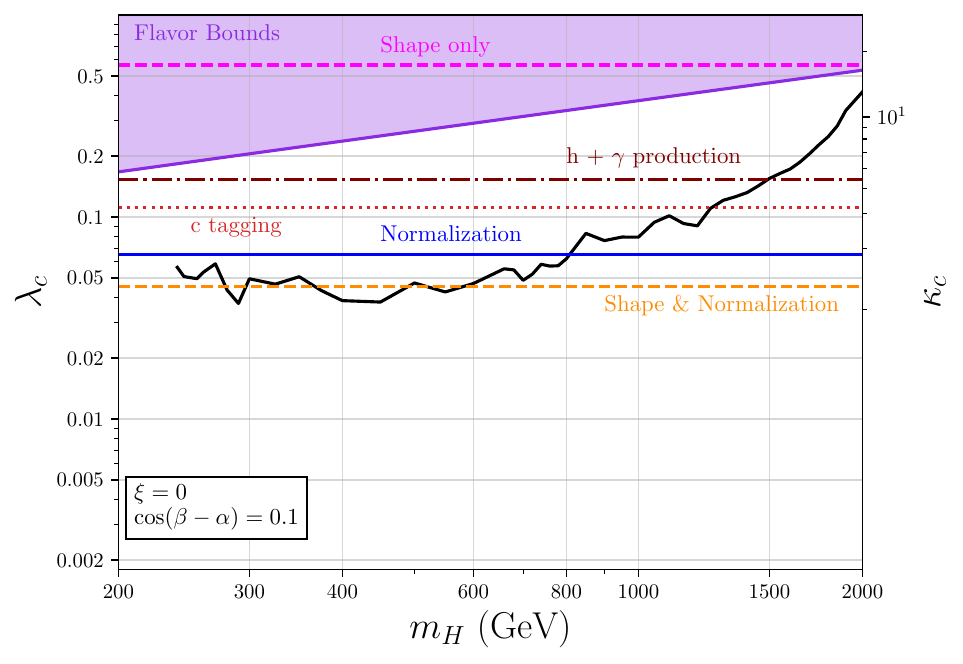} }}%
    \subfloat[]{{\includegraphics[scale=0.46]{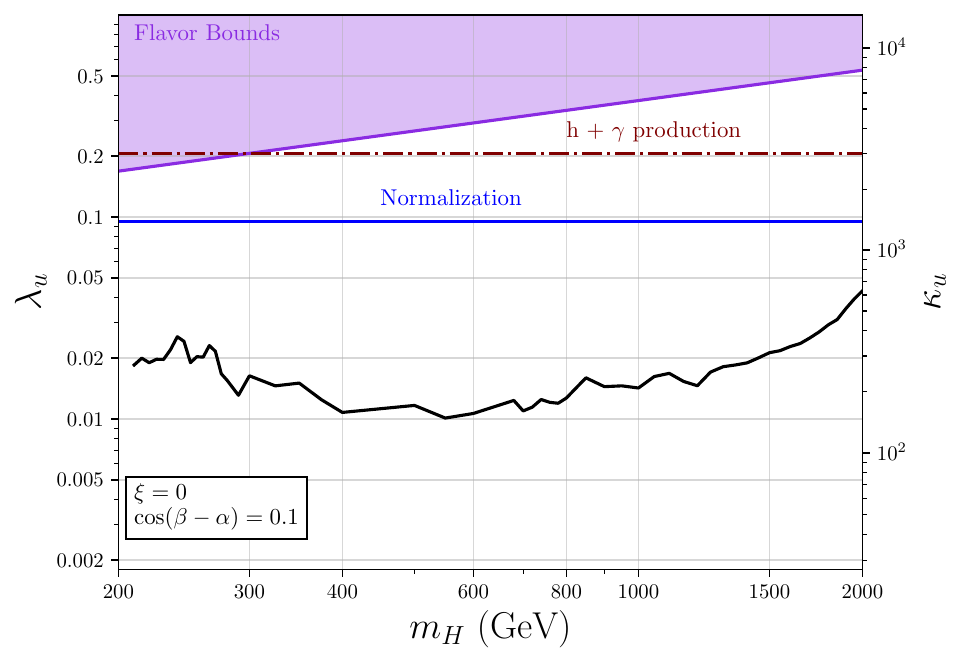} }}\\ 
    \subfloat[]{{\includegraphics[scale=0.46]{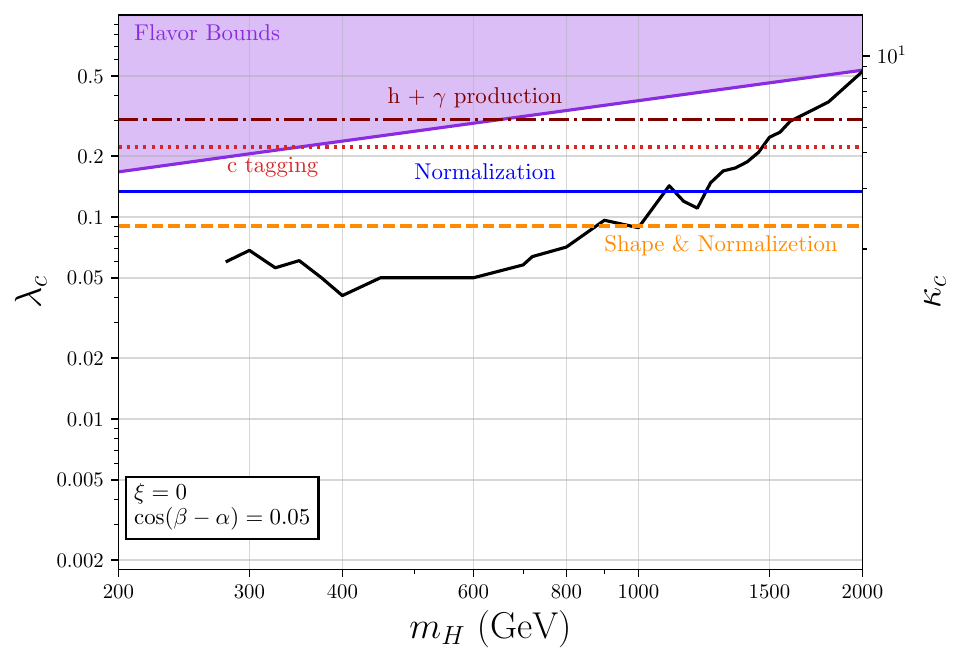} }}%
    \subfloat[]{{\includegraphics[scale=0.46]{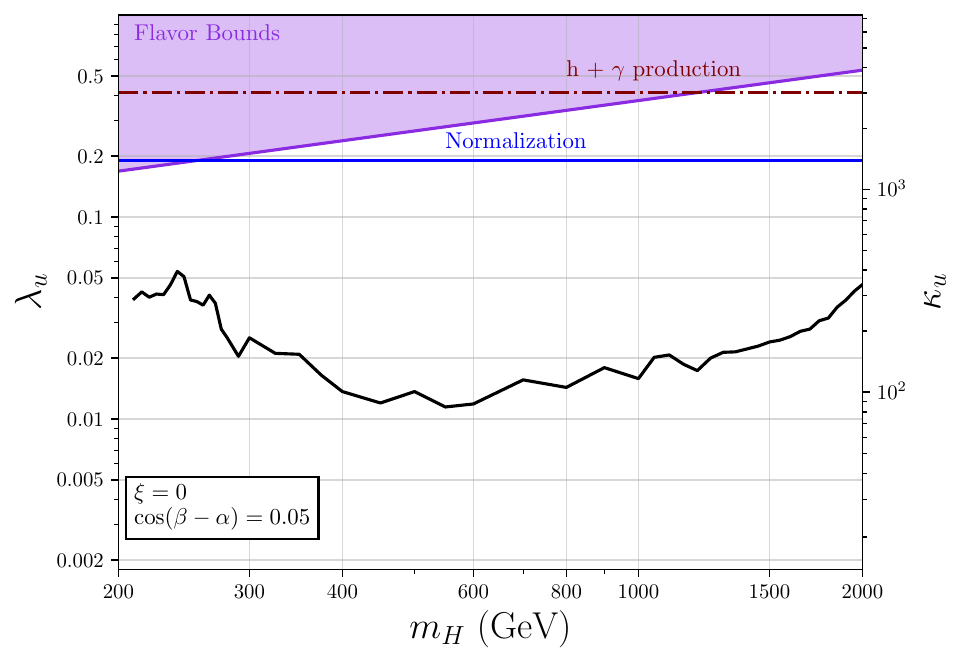} }}%
    \caption{We present the constraints for the new Yukawa couplings of our 2HDM model derived using single-Higgs properties as measured at the LHC at 95 \% C.L. The combined constraints from resonant decays of the Higgs from \autoref{fig:LHC_combined} are shown in black and overlaid with the flavor bounds from \autoref{sec:flavor_bounds}. Panel (a): Constraints of the new charm Yukawa coupling. The solid blue line shows the bounds from the Higgs signal modifier for the $\mu^{\gamma \gamma}$ channel \cite{ATLAS:10years} assuming that the central value is the SM i.e. $\mu^{\gamma \gamma}=1$. The dotted red lines correspond to the constraint obtained by c-tagging searches \cite{ATLAS-CONF-2024-010}. The dashed orange line corresponds to constraints using both shape and normalization of $\kappa_c$ derived from $p_t$ measurements \cite{ATLAS:2022fnp}. When compared to the blue normalization line, this shows that normalization dominates the results. The maroon dotted dashed line corresponds to constraints from $h + \gamma$ production \cite{CMS-PAS-HIG-23-011}. Panel (b): Constraints on the modified up Yukawa coupling. The combined constraints from resonant decays of the Higgs eigenstates from \autoref{fig:LHC_combined} are shown in black. The solid blue line corresponds to the bounds derived from the Higgs signal modifier for the $\mu^{\gamma \gamma}$ channel \cite{ATLAS:10years}. The maroon dotted dashed line corresponds to constraints from $h \gamma$ production \cite{CMS-PAS-HIG-23-011}. Panels (c) and (d) are the same as (a) and (b) respectively but for $\cos \left( \beta - \alpha \right) = 0.05$.  }%
    \label{fig:Single_higgs_xi0}%
\end{figure}


In \autoref{tab:higgs_refs} we present all the single-Higgs precision studies relevant for the modification of light quark Yukawas that we use in this study.  We give results of the studies as a single modifier of the charm Yukawa coupling, $\kappa_c$ to give an overview of the sensitivity of all charm related measurements performed thus far.  These results are not being combined, but only serve to illustrate the sensitivity to individual methods.  In most cases the experiments have made the $\kappa_c$ interpretation themselves, but for the ``normalization" of the best measured exclusive channels we have recast this ourselves, {\em assuming} a central value of the SM prediction and {\em only} a modification of the branching fractions. The purpose of this is to give an understanding of how constraining non-charm final states are given their high precision, even though they are a more indirect constraint. We have also recast the exclusive final state $h \to J/\psi Z$ from~\cite{CMS:2022fsq} using the results of~\cite{Alte:2016yuw} as this was reported only as an upper limit on the branching fraction by CMS. All the normalization bounds can also be recast trivially for a $\kappa_u$ only modification as well. However, charm tagging of the Higgs or other associated final states cannot be recast for $\kappa_u$.  One can also tag or look for exclusive mesonic decays into the first generation, and in some cases we note where an experiment has performed this study and interpretation.  However, the projected bounds on $\kappa_u$ are weaker than existing flavor and direct constraints and therefore not relevant at the LHC or HL-LHC for this class of models.
\begin{table}[h]
    \centering
    \renewcommand{\arraystretch}{1.5}
    \resizebox{\textwidth}{!}{%
    \begin{tabular}{|c|c|c|c|c|}
        \hline
         \textbf{Experiment} & \textbf{Reference} & \textbf{$\mathcal{L} (fb^{-1})$} & \textbf{Constraints} & \textbf{Comments} \\
         \hline
         ATLAS & \cite{ATLAS:10years} & 139 & $ \left| \kappa_c \right| < 2.91$ & Normalization \\
         \hline
         CMS & \cite{CMS:10years} & 138 & $\left| \kappa_c \right| < 2.92$ & Normalization \\
         \hline
         ATLAS & \cite{ATLAS:2022qef} & 139 & \makecell{$\kappa_c\in (-2.27, 2.27)$ \\ 
         $\kappa_c \in (-8.6,17.3)$} & \makecell{ \makecell{($p_T$ measurement,\\ Shape \& Normalization)} \\ $p_T$ measurement, Shape only }\\
         \hline
         CMS & \cite{CMS-PAS-HIG-23-011} & 138 & $\kappa_c \in (-6,5.4)$ & $h+\gamma$ associated production \\
         \hline
         CMS & \cite{CMS:2022psv} & 138 & $ 1.1 < \left| \kappa_c \right| < 5.5$ & c-tagging \\
         \hline
         ATLAS & \cite{ATLAS-CONF-2024-010} & 140 & $ \left| \kappa_c \right| < 4.2$ & c-tagging\\
         \hline
         CMS & \cite{CMS-PAS-HIG-23-010} & 138 & $ \left| \kappa_c \right| < 38.1$ & $h+c$ associated production\\
         \hline
         ATLAS & \cite{ATLAS:2022rej} & 139 & $\kappa_c / \kappa_{\gamma} \in (-153, 175)$ & $h \to J/\psi \gamma$ decays\\
         \hline
         CMS & \cite{CMS:2022fsq} & 138 & $\kappa_c \in (-7.5\times 10^3,7.7\times 10^3)$ & $h \to J/\psi Z$ decays \\
         \hline
    \end{tabular}%
    }
    \caption{Table of Higgs precision results, note that the results quoted
    in this table correspond to 95 \% C.L. and are recast as bounds on the charm quark coupling modifier.  In cases where the experimental result does not include an interpretation in terms of $\kappa_c$ we mark with an asterisk and refer to the text.  Using the normalization measurements we can place similar bounds for the up quark as well. Note also that in reference \cite{CMS-PAS-HIG-23-011} there is an explicit translation of their results into coupling modifier language, getting $\kappa_u \in [-3,3] \times 10^{3}$ at 95 \% C.L. For the case of \cite{ATLAS:10years, CMS:10years}, in order to translate the signal modifier constraint $\mu^{\gamma \gamma}$ to the charm coupling modifier presented in this table, we assume that the only modification to the SM couplings is the one for the charm Yukawa and that the production cross section does not get modified. For both cases we center the measurements at the SM prediction namely $\mu^{\gamma \gamma} = 1$. Finally, for the case of \cite{CMS:2022fsq}, we use the expression in \cite{Alte:2016yuw} to extract the constraint presented on the table. Notice that ATLAS has also recently studied associated production of charm and Higgs inclusively, providing a bound with no $\kappa_c$-tagging \cite{ATLAS:2024ext}. }
    \label{tab:higgs_refs}
\end{table}

In panels (a) and (b) of ~\autoref{fig:Single_higgs_xi0} we present the results of the flavor and direct searches from the previous sections overlaid with the constraints from the single-Higgs precision results given in~\autoref{tab:higgs_refs} for an alignment parameter of $\cos (\beta - \alpha )=0.1$.   In panel (a) of~\autoref{fig:Single_higgs_xi0}, the results for how charming the SM Higgs can be are shown.  Note that there is strong complementarity between the flavor, direct collider search, and precision single-Higgs results. In particular, for $m_H\lesssim 300\,\mathrm{GeV}$ and $m_H\gtrsim 1.2\,\mathrm{TeV}$, the Higgs signal strength normalization constraints are the strongest, while in the intermediate regime the direct searches provide the strongest bounds on the charm Yukawa coupling.  It is interesting to note that while charm tagging is currently a weaker bound than normalization, in the large mass regime it still provides constraints beyond the direct collider searches.  Therefore in an extension of this model where the normalization bound is ameliorated, it would be possible for charm tagging to provide valuable new information compared to other constraints.  Additionally, since charm tagging has room for substantial improvement based on methodology not just statistics, the fact that there is parameter space that is {\em already} sensitive beyond other observables, gives strong motivation to continue improving charm tagging performance. In panel (b) of~\autoref{fig:Single_higgs_xi0}, we see the analogous constraints on $\kappa_u$. There is an important distinction compared to $\kappa_c$, in all regions of parameter space the di-boson and di-Higgs searches provide the strongest constraint compared to Higgs precision. Therefore for the foreseeable future, the strongest bounds on first generation quark Yukawas don't come from Higgs precision or flavor, but by searching for the flavored progenitors at high energy.

To help illustrate the behavior for smaller values of the alignment parameter, we show the same parameter space as \autoref{fig:Single_higgs_xi0} panels (a) and (b) but for $\cos (\beta - \alpha )=0.05$  in panels (c) and (d) of \autoref{fig:Single_higgs_xi0}. Note that roughly the same conclusions can be drawn for all panels of ~\autoref{fig:Single_higgs_xi0} but the single-Higgs precision bounds are weaker due to the smaller alignment parameter for panels (c) and (d). 

\begin{figure}[h]
    \centering
    \subfloat[]{{\includegraphics[scale=0.46]{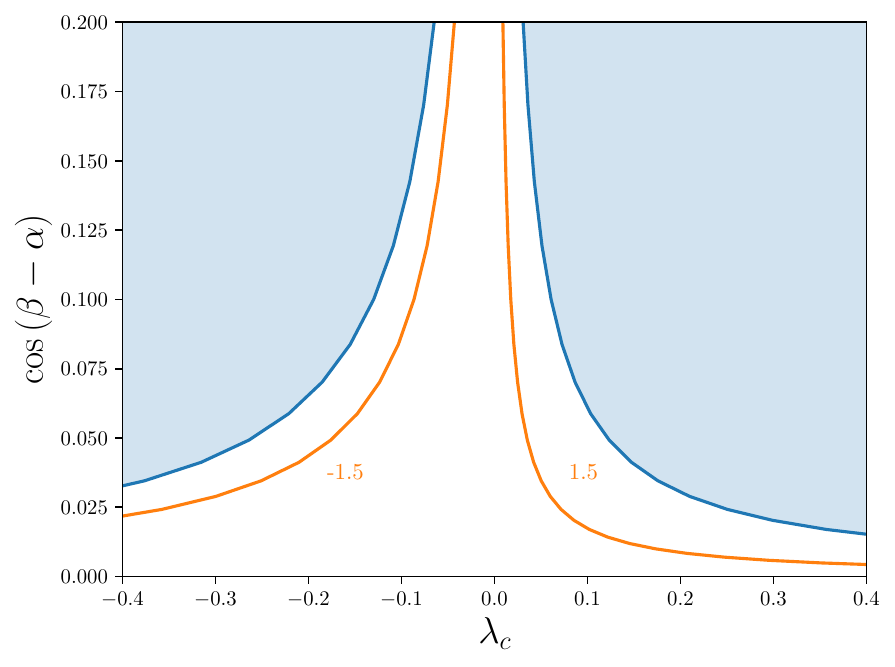} }}%
    \subfloat[]{{\includegraphics[scale=0.46]{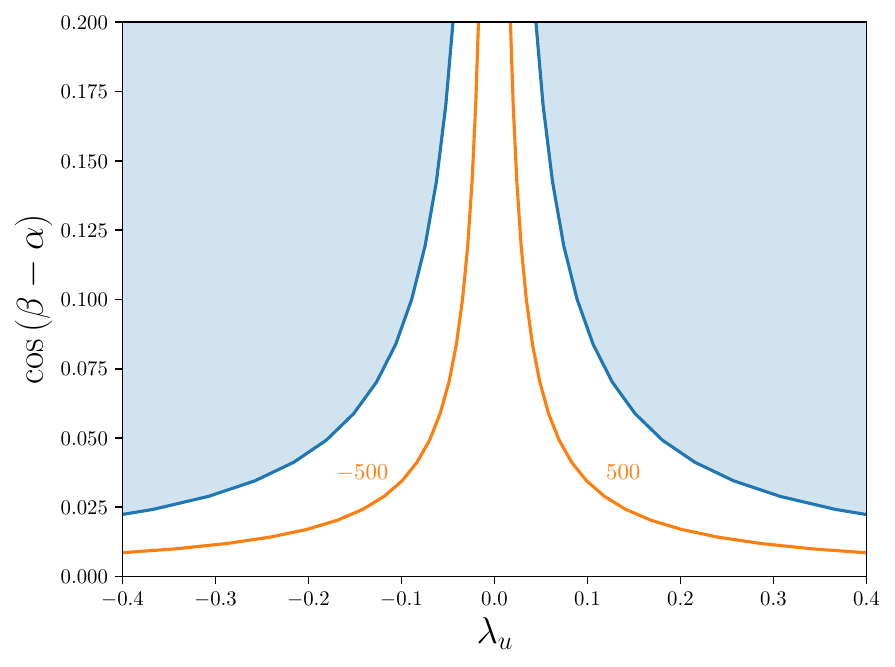} }}%
    \caption{Constraints on the allowed values of the $\lambda_c$ - $\cos (\beta -\alpha)$. In order to derive these constraints we use the signal modifier $\mu_{ggF}^{\gamma \gamma}$ from \cite{ATLAS:10years}. Assuming that only $\lambda_q$ is non-zero,  with $q = c $ for the left panel  and $q=u$ for the right one, we can extract limits for the coupling modifier $\kappa_{q_i}$ at 95 \% C.L. which we can then recast in the parameter space of our model, shown as the shaded blue region. Note that for the $\mu_{ggF}^{\gamma \gamma}$ to $\kappa_{q_i}$ conversion, we also assume that the production cross sections do not change from their SM values. For the case of $\lambda_u$ the results are almost symmetric around $\lambda_u = 0 $ because $\kappa_u$ is significantly less constraint compared to $\kappa_c$ and dominates the expression.  Finally, in order to make comparison with the SM easier, we present the contour lines for $\kappa_u = \pm 500$ and $\kappa_c = \pm 1.5$ within the model.}
    \label{fig:coupling_modifier}
\end{figure}

Finally, to understand more generally the interplay with precision Higgs constraints, we show the allowed region as function of the new Yukawa couplings $\lambda_u$ and $\lambda_c$ and  $\cos(\beta-\alpha)$ in \autoref{fig:coupling_modifier}. We also plot contours of constant $\kappa_u$ and $\kappa_c$ to demonstrate the large deviations possible even at extremely small values of $\cos(\beta-\alpha)$. 

\section{Conclusions}
\label{sec:concl}
\qquad In this work we have focused on the phenomenology of a 2HDM realizing down-type Spontaneous Flavor Violation (SFV).  This allowed for a concrete realization of models which could significantly modify the {\em charm} and/or {\em up} quark Yukawa couplings of the SM Higgs while maintaining consistency with all precision flavor experiments. 

In particular, we have demonstrated that flavor constraints can allow for a second Higgs with Yukawa couplings of $\mathcal{O}(10^{-1})$ to any up-type quark {\em without} having to respect any of the hierarchies of the SM Yukawas.  For example one could have a new Higgs that couples at this strength {\em only} to the charm quark.  When this new Higgs has a non-zero mixing with the SM Higgs, this allows for large deviations of $\mathcal{O}(10)$ to $\mathcal{O}(10^4)$ to the SM charm and up Yukawa couplings consistent with measurements of FCNCs.

Flavor constraints are not the only relevant constraints for a new state near the EW scale with potentially large couplings to light quarks. Direct search constraints are particularly powerful given that if there is mixing between the SM Higgs and heavier Higgs this implies that the heavier Higgs can decay into di-bosons and di-Higgs. We find that these direct search constraints are always stronger than flavor bounds in the limit of small but non-zero mixing for the down-type SFV 2HDM cases we explored.

Finally, we examined whether or not the allowed parameter space was consistent with the Higgs precision measurements.  The strongest constraints from single-Higgs precision currently come from the best measured exclusive channels.  However, we have shown that the direct search constraints for the new Higgs states which could cause measurable SM Higgs deviations are generically more powerful at the LHC.  At very low or high masses of the second Higgs state the indirect single-Higgs precision constraints are stronger for charm Yukawa deviations, while for the up-quark case we found direct search constraints and flavor are always stronger. We have also updated the up-type SFV 2HDM bounds from~\cite{Egana-Ugrinovic:2021uew} in Appendix~\ref{sec:updated_up_type}, which reaches a similar conclusion. While indirect single-Higgs precision constraints are powerful, what is more exciting is the possibility of improvements in direct charm-tagging measurements at the LHC and beyond in light of the extraordinary advancement provided by machine learning in High Energy.  We have demonstrated that there are viable models consistent with flavor and direct searches that could be explored in a new way with improvements in charm tagging.  This is of particular significance as current projections require an $\mathcal{O}(1)$ deviation in the charm Yukawa for LHC sensitivity which is difficult to account for in standard flavor models. 

Finally it is important to remember that in the pursuit of how charming the Higgs can be, as well as the deviations in other light quark Yukawas, it is always tied to the UV model.  While there could be large deviations as no one has any idea of the origin of flavor in our universe, there still must be a mechanism to explain significant deviations in  while maintaining consistency with experimental results.  For any light quark Yukawa deviation that is sizable enough to be relevant at the LHC, the scale of new physics cannot be parametrically separated from the EW scale without running into unitarity bounds.   
Therefore any observables attempting to test Yukawa couplings in the SM must be compared to other data which explores possible UV completions, of which few are currently known.   Therefore the LHC is currently ideally suited to carry out this program of both precision and energy frontier measurements.  Furthermore, a future high energy $e^+e^-$ linear collider, proton collider or muon collider at the 10 TeV pcm scale would likely be able to cover the entire parameter space of known UV origins for measurable light quark Yukawa coupling deviations.  We leave investigating these options and the larger parameter space that could be explored within SFV 2HDMs for future work.

\acknowledgments
We would like to thank Hannah Arnold, Livia Soffi and Samuel Homiller for useful discussions and feedback. M.V. acknowledges support from the projects ``Exploring New Physics'' and ``Theoretical Particle Physics and Cosmology'' funded by INFN and thanks YITP and SCGP for the hospitality during the completion of the work.  The work of A.S.G. and P.M is supported in part by the National Science Foundation grant PHY-2210533.

\appendix
\section{2HDM: The up-type SFV case}
\label{sec:updated_up_type}

\qquad In this section we present the constraints we get following the up-type SFV flavor prescription for a 2HDM, updating the ones set in \cite{Egana-Ugrinovic:2019dqu}. The new results we present here are produced using new constraints on dimension-six operators and LHC data from Run 2, see references in \autoref{tab:flavor_summary} and \autoref{tab:collider_refs}. In order to make the comparison with \cite{Egana-Ugrinovic:2019dqu} straightforward, we make the same choice for $\xi$ and the alignment parameter $\cos \left( \beta -\alpha \right)$ for the flavor and collider bounds respectively. As we saw in \autoref{sec:The model}, within the SFV prescription, only one quark sector can be aligned. For the up-type SFV 2HDM, the new Yukawa couplings will appear in the down quark sector and the couplings for the second Higgs doublet will have the following form, 

\begin{equation}
    \mathcal{Y}_2^u =  \xi V^* Y^u , \quad \mathcal{Y}_2^{d } = - {\rm diag}\left( \lambda_d, \lambda_s, \lambda_b \right),  \quad  \mathcal{Y}_2^{\ell} = - \xi^{\ell} Y^{\ell} .
\label{eq:YukawasH2_uptype}
\end{equation}

Similar to the case for the down-type SFV prescription, the same parameters $\xi$ and $\xi^{\ell}$ control the coupling strength of the second doublet to the up-type quarks and the leptons, respectively. As expected, our flavor bounds will depend on the values of these two parameters. In accordance with the rest of this work, we will set $\xi^{\ell}=0$, concentrating our attention on the quark sector. Below we present the bounds derived using recent results from Flavor Physics for the allowed values of the relevant Wilson coefficients, see \autoref{tab:flavor_summary}. The functional dependence of the Wilson coefficients to the up-type SFV model parameters can be derived using the same expressions mentioned in \autoref{sec:flavor_bounds} and the up-type couplings in the Appendix of \cite{Egana-Ugrinovic:2019dqu}. Note that the resulting dependence of the coefficients on the 2HDM parameters will be different than the one for the down-type prescription. We can verify this by comparing the expressions for the up-type couplings in equation (\ref{eq:YukawasH2_uptype}) with the equivalent ones in the down-type case listed in Appendix \ref{couplings}. 

\begin{figure}[b!]
    \centering
    \includegraphics[scale=0.5]{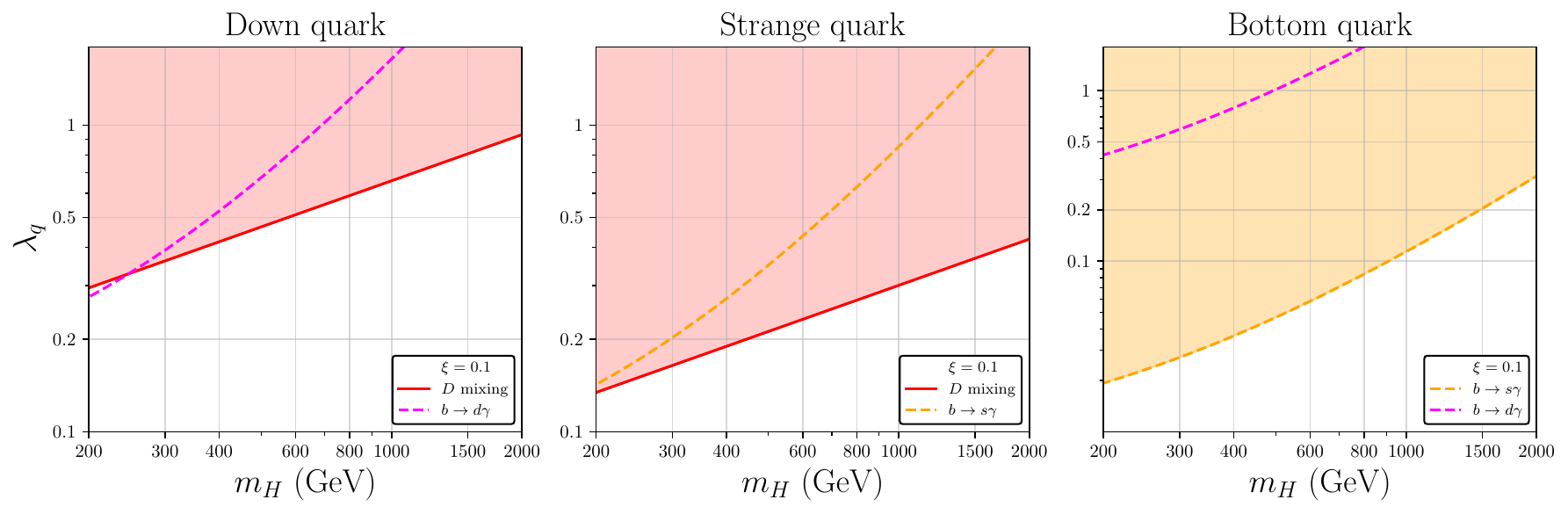}
    \includegraphics[scale=0.5]{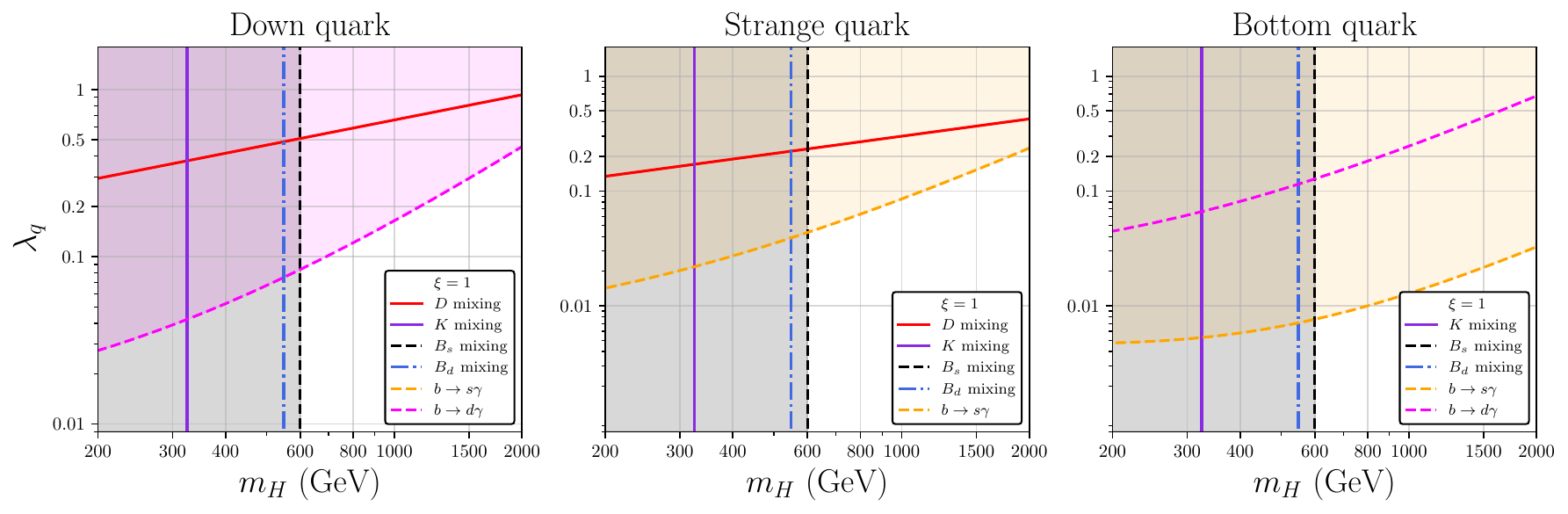}
    \caption{Bounds from FCNCs for the up-type SFV. The top row shows bounds for different couplings $\lambda_{q_i}$ with $\xi=0.1$: (left) $\lambda_d$ with $\lambda_s=\lambda_b=0$;  (middle) $\lambda_s$ with $\lambda_d=\lambda_b=0$; (right) $\lambda_b$ with $\lambda_d=\lambda_s=0$. The bottom row presents the same bounds for $\xi=1$. The value of $\cos (\beta -\alpha)$ does not affect these results since only the charged Higgs eigenstates are involved in these processes. Due to the different structure of the Yukawa matrices for the up-type SFV, we see that bounds coming from D meson mixing now become much stricter compared to the down-type SFV case shown in \autoref{fig:combined_flavor}. We present in solid red lines the constraints coming from D meson mixing and in dashed orange and magenta lines the ones coming from $b \to s \gamma$ and $b \to d \gamma$ decays respectively. The bounds of the relevant effective operators are shown in \autoref{tab:flavor_summary}. Notice that the flavor bounds depend strongly on the value of $\xi$ that we choose.}%
    \label{fig:combined_flavor_uptype_0p1}%
\end{figure}

\begin{figure}[t!]
    \centering
    \includegraphics[scale=0.6]{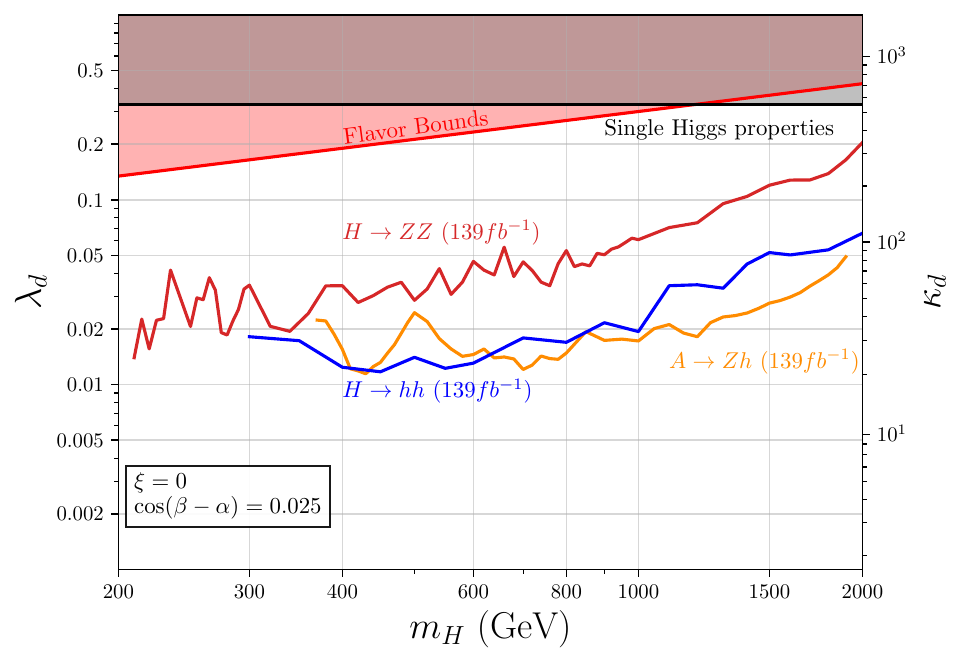}
    \caption{Collider searches for the 2HDM model with couplings to down-type quarks, overlaid with the flavor bounds for $\xi = 0$. The couplings of the second doublet to the up-type quarks are set to zero i.e. $\xi = 0$ and the alignment parameters is set to $\cos \left( \beta - \alpha \right) = 0.025$. The vertical axis on the right hand side shows the enhancement of the coupling of h to the down quarks with respect to the Standard Model Yukawa. The flavor bounds combined are presented in light red. We present the bounds of resonant production of the heavy Higgs H, decaying to different channels. The bounds from decays to a pair of Z bosons \cite{ATLAS:H-to-ZZ} are in red and to a pair of $125 \, \GeV$ Higgs \cite{ATLAS:dihiggs-bb, ATLAS:dihiggs-tata, ATLAS:dihiggs-gaga} are in blue. In orange, we present the bounds from resonant production of the CP odd A decaying to a Z boson and a $125 \, \GeV$ Higgs boson \cite{ATLAS:AZh}. Finally, in black we present the bound that we obtain from the Higgs signal modifier \cite{ATLAS:10years, CMS:10years}. }
    \label{fig:LHC_uptype}
\end{figure}

By comparing the results in the first row of  \autoref{fig:combined_flavor_uptype_0p1}, where we have set $\xi=0.1$, with the ones in the second row of the same figure, where $\xi=1$, we can verify that the Flavor Physics constraints are sensitive to the value of $\xi$. This behavior is the result of the explicit dependence of the Wilson coefficients on the NP parameters. Overall, the ability to place bounds on the parameter space of our 2HDM depends on the flavor prescription that we follow (up or down-type SFV) and the specific value we choose for $\xi$. This sensitivity depends on which quark sector contains the new couplings. With the SFV prescription we can only produce new aligned couplings for either the up or down quarks. Finally, we note that as is the case for down-type SFV, the Flavor Physics bounds for the up-type SFV 2HDM do not depend on the alignment parameter because the Higgs eigenstates involved in these processes are the charged ones. 
Finally, using \textsc{MadGraph} \cite{Madgraph5} we recast experimental bounds from collider searches in the up-type SFV model parameter space. As we did in \autoref{sec:ColliderBounds}, we concentrate on the quark sector by setting the couplings to the leptons to zero, $\xi^{\ell} = 0$. We choose to consider a regime away from alignment, $\cos \left( \beta - \alpha \right) \neq 0$. As a result, setting $\xi = 0 $ does not prevent the modification of all the $125 \, \GeV$ Higgs Yukawa couplings. The results are presented in \autoref{fig:LHC_uptype}, showing the same search channels as in \autoref{fig:LHC_combined} for the down-type SFV. 

\section{Loop functions}
\label{loop_functions}
The loop functions used in the previous sections are defined as follows \cite{Crivellin:2HDM, Crivellin:bsmumu}:
\begin{align*}
    \mathcal{C}^0_{7,XY}(z_j) &= \frac{z_j}{12} \left[ \frac{-5z_j^2+8z_j-3+(6z_j-4) \log z_j}{(z_j -1 )^3} \right] \ , \\
    \mathcal{C}^0_{8,XY}(z_j) &= \frac{z_j}{4} \left[ \frac{-z_j^2 + 4 z_j -3 -2 \log z_j}{(z_j-1)^3} \right] ,  \\
    \mathcal{C}^0_{7,YY}(z_j) &= \frac{z_j}{72} \left[ \frac{-8z_j^3 +3z_j^2 +12z_j -7 + (18z_j^2 -12z_j) \log z_j}{ (z_j -1)^4} \right] \ , \\
    \mathcal{C}^0_{8,YY}(z_j) &= \frac{z_j}{24} \left[ \frac{-z_j^3 + 6 z_j^2 - 3 z_j - 2 - 6z_j \log z_j}{(z_j -1 )^4} \right],  \\
    I_1(z_j) &= -\frac{1}{z_j-1} + \frac{\log (z_j) z_j}{(z_j-1)^2} \ .
\end{align*}
With $d=4-2\epsilon$, we get for the loop functions appearing in the box diagram calculations in eqs. (\ref{boxes_EW}) and (\ref{boxes_Higgs}):
\begin{align*}
    \mathcal{C}_0 (m_1^2, m_2^2, m_3^2) &= \frac{m_1^2 m_2^2 \log \left( \frac{m_1^2}{m_2^2}\right) + m_3^2 m_2^2 \log \left( \frac{m_2^2}{m_3^2}\right) + m_3^2 m_1^2 \log \left( \frac{m_3^2}{m_1^2}\right)}{(m_1^2 -m_2^2)(m_3^2 - m_1^2)(m_2^2 - m_3^2)} \ , \\
    D_0 (m_1^2, m_2^2, m_3^2, m_4^2) &= \frac{\mathcal{C}_0 \left( m_1^2, m_2^2, m_3^2 \right) - \mathcal{C}_0 \left(m_1^2, m_2^2, m_4^2 \right)}{m_3^2 - m_4^2} \ , \\
    D_2 (m_1^2, m_2^2, m_3^2, m_4^2) &= \mathcal{C}_0 \left( m_1^2, m_2^2, m_3^2 \right) + m_4^2 D_0 (m_1^2, m_2^2, m_3^2, m_4^2) \ .
\end{align*}

\section{Couplings in down-type SFV}
\label{couplings}
In \autoref{tab:couplings} we show the physical couplings of the Higgs eigenstates to the fermions derived from Eq.~\eqref{eq:HiggsLagrangian} for the down-type SFV case. In order to keep the notation as clear as possible, in the table we denote by bar the right handed fields. We denote by $Y^q$ with $q=u,d, \ell$ the Standard Model Yukawa matrices for up, down-type quarks and leptons respectively. $\Lambda^u = {\rm diag} \left( \lambda_u, \lambda_c, \lambda_t \right)$ is the diagonal matrix containing the three new couplings introduced in the down-type SFV 2HDM. Couplings are defined with a negative sign in the Lagrangian, i.e. $\mathcal{L} \supset - \lambda_{hf\overline{f}} h f \overline{f}$.
   
\begin{table}[h]
    \centering
    \resizebox{\textwidth}{!}{%
    \begin{tabular}{|c|c|c|c|}
        \hline
        $\mathcal{Y}^h_{u_i\overline{u}_j}$ & $\delta_{ij} \left[ Y^u_i \sin \left( \beta - \alpha \right) + \Lambda_i^u \cos \left( \beta - \alpha \right) \right]$ &  $\mathcal{Y}^H_{ u_i \overline{u}_j}$ & $\delta_{ij} \left[ -Y^u_i \cos \left( \beta - \alpha \right) + \Lambda_i^u \sin \left( \beta - \alpha \right) \right]$ \\ 
        $\mathcal{Y}^h_{ d_i \overline{d}_j}$ & $\delta_{ij} Y^d_i \left[ \sin \left( \beta - \alpha \right) + \xi \cos \left( \beta - \alpha \right) \right]$ & $\mathcal{Y}^H_{d_i \overline{d}_j} $ & $\delta_{ij} Y^d_i \left[ - \cos \left( \beta - \alpha \right) + \xi \sin \left( \beta - \alpha \right) \right]$\\
        $\mathcal{Y}^h_{ \ell_i \overline{\ell}_j} $ & $\delta_{ij}  Y^{\ell}_i \left[ \sin \left( \beta - \alpha \right) + \xi^{\ell} \cos \left( \beta - \alpha \right) \right]$  & $\mathcal{Y}^H_{ \ell_i \overline{\ell}_j}$ & $\delta_{ij} Y^{\ell}_i \left[ - \cos \left( \beta - \alpha \right) + \xi^{\ell} \sin \left( \beta - \alpha \right) \right]$\\
        $\mathcal{Y}^A_{ u_i \overline{u}_j} $ & $i \delta_{ij} \Lambda_i^u$ & $\mathcal{Y}^{H^+}_{ d_i \overline{u}_j}$ & $ \left( V^T \Lambda^u \right)_{ij}$\\
        $\mathcal{Y}^A_{d_i \overline{d}_j}$ & $-i\xi \delta_{ij} Y_i^d$ & $\mathcal{Y}^{H^-}_{u_i \overline{d}_j}$ & $-\left( \xi V^* Y^d \right)_{ij}$\\
        $\mathcal{Y}^A_{ \ell_i \overline{\ell}_j}$ & $-i \xi^{\ell} \delta_{ij} Y_i^{\ell}$ & $\mathcal{Y}^{H^-}_{ \ell_i \overline{\ell}_j}$ & $- \left( \xi^{\ell} Y^{\ell} \right)_{ij}$ \\
        \hline
    \end{tabular}}
    \caption{Higgs couplings to the fermions in the mass basis, see the text for details.}
    \label{tab:couplings}
\end{table}


\bibliographystyle{JHEP-CONF}
\bibliography{biblio.bib}

\end{document}